\documentclass[aps,prb,twocolumn,showpacs,reprint]{revtex4-1}
\usepackage{hyperref}
\usepackage{amsfonts}
\usepackage{amssymb}
\usepackage{amsmath}
\usepackage{graphicx}
\usepackage[utf8]{inputenc}
\usepackage{subfigure}

\begin{document}

\title{Fermionic two-loop functional renormalization group for correlated
fermions: Method and application to the attractive Hubbard model}

\date{\today}

\author{Andreas Eberlein}

\affiliation{Max Planck Institute for Solid State Research, 
 D-70569 Stuttgart, Germany}


\newcommand{\etal}{\textit{et~al.}}
\newcommand{\ie}{\textit{i.\,e.}}

\newcommand{\tr}{\operatorname{tr}}

\newcommand{\V}[1]{\Gamma^{(4)\Lambda}_{#1}}

\newcommand{\VPH}[1]{V^{\text{PH},\Lambda}_{#1}}
\newcommand{\VPP}[1]{V^{\text{PP},\Lambda}_{#1}}

\newcommand{\dVPHdL}[1]{\dot V^{\text{PH},\Lambda}_{#1}}
\newcommand{\dVPPdL}[1]{\dot V^{\text{PP},\Lambda}_{#1}}

\newcommand{\G}[1]{G^\Lambda_{#1}}
\newcommand{\SL}[1]{S^\Lambda_{#1}}

\newcommand{\mtau}[2]{\tau^{(#1)}_{#2}}

\newcommand{\intdrei}[1]{\int\negthickspace\negthinspace\frac{d^3 #1}{(2\pi)^3}}


\hyphenation{counter-term}

\begin{abstract}
We derive an efficient method for treating renormalization contributions at
two-loop level within the functional renormalization group in the one-particle
irreducible formalism for fermions. It is based on a decomposition of the 
two-particle vertex in charge, magnetic and pairing channels. The method treats
single-particle and collective excitations in all channels on equal footing,
allows for the description of symmetry-breaking and captures collective mode 
fluctuation physics in the infrared. As a first application, we study the 
superfluid ground state of the two-dimensional attractive Hubbard model. We 
obtain superfluid gaps that are reduced by fluctuations in comparison to the 
one-loop approximation and demonstrate that the method captures
the renormalization of the amplitude mode by long-range phase fluctuations.
\end{abstract}

\maketitle

\section{Introduction}
In the last decade, the functional renormalization group (fRG) proved itself as
an excellent tool for the study of competing instabilities in correlated
electron systems and a valuable source of new approximation schemes.\cite{*[{For
a review, see }] Metzner2012} It treats charge, magnetic and pairing channels on
equal footing and in a scale-separated way. This allows for the study of
competing order and fluctuation driven instabilities. One of the major successes
of fRG was to provide evidence for $d$-wave superconductivity in the
two-dimensional repulsive Hubbard model at weak and intermediate
couplings.\cite{Zanchi2000,Halboth2000a,Honerkamp2001a,Eberlein2014}
The method was also successfully applied to quantum wires and dots in and out of
equilibrium,\cite{Andergassen2004,Karrasch2010} multiband systems describing
pnictide superconductors~\cite{Wang2009a,Thomale2009,Platt2009} and also spin
models.\cite{Reuther2010,Reuther2011}

Most of these fRG studies used the one-particle irreducible (1PI) formalism
in a purely fermionic formulation and were restricted to the one-loop level
without or with taking the self-energy feedback into account in the flow
equation for the vertex as suggested by Katanin.\cite{Katanin2004a} The latter
modification improves the fulfillment of Ward identities and allows to continue
fermionic fRG flows into phases with broken
symmetries.\cite{Salmhofer2004,Gersch2005,Gersch2008,Maier2012,Eberlein2013,
Eberlein2014,Maier2014}
However, in a recent study of superfluidity in the attractive
Hubbard model, it was pointed out that the singular infrared behavior of 
the amplitude mode of the gap due to phase 
fluctuations~\cite{Castellani1997a,Pistolesi2004} is not captured in
the Katanin scheme.\cite{Eberlein2013,Eberlein2013-thesis} 

An alternative route to studying fluctuation physics and symmetry breaking is
to decouple the fermionic fields by introducing bosonic auxiliary fields via a 
Hubbard-Stratonovich transformation. Applying the fRG to the resulting mixed 
fermion-boson action captures the asymptotic infrared behavior of fermionic 
superfluids already at one-loop level within a relatively simple 
truncation.\cite{Strack2008,Obert2013} Besides fermionic 
superfluidity,\cite{Birse2005,Strack2008,Obert2013} this route was also 
pursued for studying the BEC-BCS crossover in continuum 
systems,\cite{Diehl2007,Bartosch2009,Diehl2010}
and antiferromagnetism~\cite{Baier2004} as well as $d$-wave
superconductivity~\cite{Friederich2011} in the two-dimensional Hubbard model.

The choice of auxiliary fields introduces some bias if only a small number of
them and simple truncations are used.\cite{Baier2004} This bias
can be reduced by dynamical rebosonization.\cite{Gies2002,Floerchinger2009} Most 
fRG studies of partially bosonized actions relied on relatively simple ansätze 
for the bosonic potential. However, in situations with competition of 
instabilities, truncating the order parameter potential is not straightforward. 
It may therefore be advantageous to treat at least the high and intermediate 
scales in a purely fermionic language. In addition, results for the vertex in 
the Hubbard model at large couplings indicate that the effective interaction 
contains contributions that are not well described as boson-mediated
interactions.\cite{Rohringer2012,Kinza2013} Recently, Veschgini and Salmhofer
derived a hierarchy of flow equations for fermionic 1PI vertex functions from
Dyson-Schwinger equations and pointed out that including all renormalization
contributions at two-loop level may increase the degree of
self-consistency.\cite{Veschgini2013}

All this motivates going beyond the one-loop approximation or Katanin's scheme
within the purely fermionic formalism. One fRG study at two-loop level has
been carried out for the repulsive Hubbard model in the symmetric regime by
Katanin.\cite{Katanin2009} He used the $N$-patch approximation for the
momentum dependence of the vertex and neglected its frequency dependence. For
relatively weak couplings, he concluded that fluctuations at two-loop level do 
not change the flow qualitatively. This is not clear at larger couplings and in 
symmetry-broken phases. Besides, in such situations the frequency dependence of 
the vertex should be taken into account, which seems to be beyond the scope of 
Katanin's two-loop approach due to the resulting numerical complexity.

In this article, we present a reformulation of the two-loop flow equations that
is effectively of one-loop form. It is exact to the third order in the
effective interaction and based on a decomposition of the vertex in charge,
magnetic and pairing channels.\cite{Karrasch2008,Husemann2009} It
allows to take the frequency dependence of the vertex into account with a
reasonable numerical effort and to continue fRG flows into symmetry-broken
phases. We demonstrate that the method captures the singular asymptotic
infrared behavior of a fermionic $s$-wave superfluid. Besides the general
scheme, we present as a first application a study of the superfluid ground state
of the attractive Hubbard model. In the presence of a not too small external
pairing field, the one- and two-loop flows are qualitatively similar, albeit
with smaller critical scales and gaps at two-loop level. The infrared
behavior is different in the two approximations, as expected.

The article is structured as follows. In Sec.~\ref{sec:Method}, we describe
the reformulated flow equations at two-loop level that allow for an
efficient numerical solution. In Sec.~\ref{sec:AHM}, results for
the attractive Hubbard model are presented, including estimates for the
infrared behavior and numerical results. Section~\ref{sec:Summary}
contains a short summary and final remarks.

\section{Method}
\label{sec:Method}
In this section, we describe the reformulation of the two-loop flow equations
as effective one-loop equations, which is at the heart of this article. The
result is applicable to all systems whose effective actions contain vertices
with an equal number of creation and annihilation operators only (see below) and
in which it is meaningful to classify diagrams according to singular dependences
on transfer momenta or frequencies.

A general derivation of renormalization group equations in the 1PI framework can
be found for example in Ref.~\onlinecite{Metzner2012}. Flow equations at 
two-loop level were derived for example in
Refs.~\onlinecite{Salmhofer2004,Veschgini2013}.

\subsection{Flow equations at two-loop level}
\label{subsec:TL_FlowEq}
In this section, we outline the derivation of flow equations at two-loop 
level in order to make the article self-contained, following the presentations
in Refs.~\onlinecite{Salmhofer2004,Eberlein2013-thesis}. 

Flow equations at two-loop level are based on a truncation of the flowing
effective action $\Gamma^\Lambda[\bar\phi,\phi]$ at the three-particle level,
\begin{equation}
\begin{split}
	\Gamma^\Lambda&[\bar\phi,\phi] = \Gamma^{(0)\Lambda} + \sum_{\alpha,\beta}
\Gamma^{(2)\Lambda}_{\alpha\beta} \bar\phi_\alpha \phi_\beta \\
	& + \frac{1}{4} \sum_{\alpha,\beta,\gamma,\delta} \V{\alpha \beta \gamma
\delta} \bar\phi_\alpha \bar\phi_\beta \phi_\gamma \phi_\delta\\
 & +\frac{1}{(3!)^2}\sum_{\alpha,\beta,\gamma,\delta,\epsilon,\zeta}
\Gamma^{(6)\Lambda}_{\alpha\beta\gamma\delta\epsilon\zeta}
\bar\phi_\alpha\bar\phi_\beta
\bar\phi_\gamma \phi_\delta \phi_\epsilon \phi_\zeta + \ldots,
\label{eq:AnsatzEffAction}
\end{split}
\end{equation}
where $\bar\phi$, $\phi$ are anticommuting Grassmann fields and
$\Gamma^{(2n)\Lambda}$ the 1PI $n$-particle vertex
functions\footnote{$\Gamma^{(0)\Lambda}$ describes the interaction
correction to the grand-canonical potential and is not important in the
following.}. The
ellipsis represents four-particle and higher-order terms. The Greek indices
$\alpha = (k, s)$, with $k = (k_0, \boldsymbol k)$, collect momenta
$\boldsymbol k$ as well as fermionic Matsubara frequencies $k_0$ and the spin or
Nambu index $s$. For describing a system in the symmetric phase, the fermionic
fields can be chosen in spinor representation, $\phi_\alpha \rightarrow \psi_{k
s}$ and $\bar\phi_\alpha \rightarrow \bar\psi_{k s}$, where $s
=\uparrow\downarrow$ is the spin orientation. For a fermionic singlet 
superfluid or superconductor, as considered in the following, the fields are 
most conveniently chosen in \emph{Nambu} representation, where $s = \pm$ is the 
Nambu index and
\begin{equation} \label{eq:NambuRepresentation}
 \phi_{k+} = \psi_{k\uparrow}, \quad 
 \bar\phi_{k+}  = \bar\psi_{k\uparrow}, \quad
 \phi_{k-} = \bar\psi_{-k\downarrow}, \quad 
 \bar\phi_{k-}  = \psi_{-k\downarrow}.
\end{equation}
Using this representation in the presence of spin rotation invariance, only
vertices that create and annihilate an equal number of Nambu quasiparticles
appear in the effective action.\cite{Gersch2008,Eberlein2010}

\begin{figure}
	\centering
	\includegraphics[scale=.75]{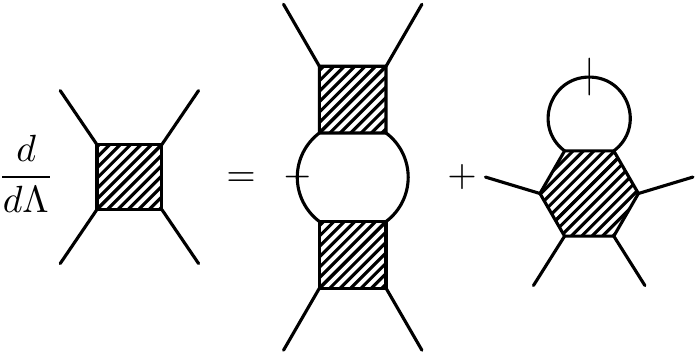}
	\caption{Simplified diagrammatic representation of the one-loop
renormalization group equation for the two-particle vertex.}
	\label{fig1:FlowEq2P}
\end{figure}

Inserting the ansatz for the effective action Eq.~\eqref{eq:AnsatzEffAction}
into the functional flow equation and evaluating appropriate functional
derivatives with respect to the fields (see Ref.~\onlinecite{Metzner2012}) 
yields flow equations for the self-energy
\begin{equation}
	\frac{d}{d\Lambda} \Sigma^\Lambda_{\alpha\beta} = \sum_{\gamma,\delta}
\SL{\delta\gamma} \V{\alpha\gamma\delta\beta}
\end{equation}
and the two-particle vertex
\begin{equation}
\begin{split}
	\frac{d}{d\Lambda} &\V{\alpha\beta\gamma\delta} = \\
	& \sum_{a,b,c,d}
\Bigr[\partial_{\Lambda,S}(\G{ab} \G{cd}) (\V{\alpha b c
\delta} \V{d \beta\gamma a} - \V{\beta b c \delta} \V{d\alpha\gamma a}) \\
	& -\frac{1}{2} \partial_{\Lambda,S}(\G{ab}\G{cd}) \V{\alpha\beta c a} \V{b d
\gamma \delta}\Bigr] + \sum_{a,b} \SL{ba} \Gamma^{(6)\Lambda}_{\alpha\beta a b
\gamma \delta},
\label{eq:fRG:dVdL}
\end{split}
\end{equation}
where $\G{\alpha\beta} = (\Gamma^{(2)\Lambda})^{-1}_{\alpha\beta}$ is the full
fermionic propagator and $\SL{\alpha\beta} = (\frac{d}{d\Lambda}
\G{\alpha\beta})_{\Sigma^\Lambda = \text{const.}} \equiv \partial_{\Lambda,S}
\G{\alpha\beta}$ the so-called single-scale propagator. The self-energy is
connected to $\Gamma^{(2)\Lambda}$ via a Dyson-Schwinger equation,
$(\G{})^{-1}_{\alpha\beta} = (G^\Lambda_{0})^{-1}_{\alpha\beta} -
\Sigma^\Lambda_{\alpha\beta}$, where $G^\Lambda_0$ is the regularized bare
propagator at scale $\Lambda$. The flow equation for the two-particle
vertex is illustrated diagrammatically in Fig.~\ref{fig1:FlowEq2P}.

\begin{figure}
	\centering
	\includegraphics[scale=.75]{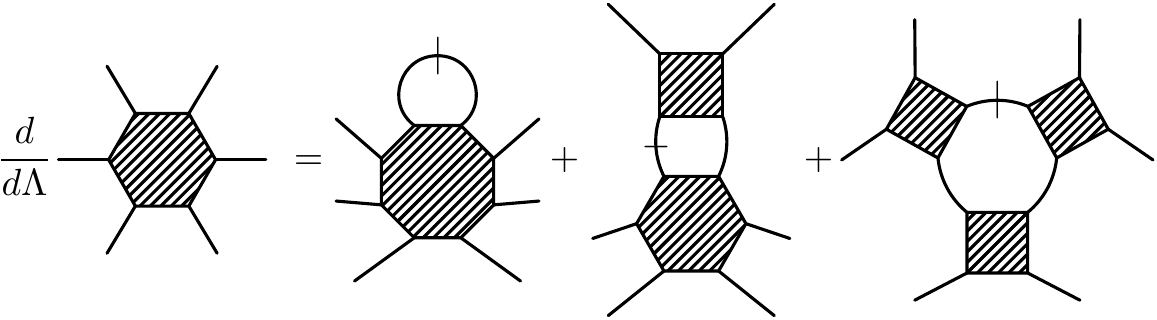}
	\caption{Simplified diagrammatic representation of the one-loop
renormalization group equation for the three-particle vertex.}
	\label{fig2:FlowEq3P}
\end{figure}
\begin{figure}
	\centering
	\includegraphics[scale=.75]{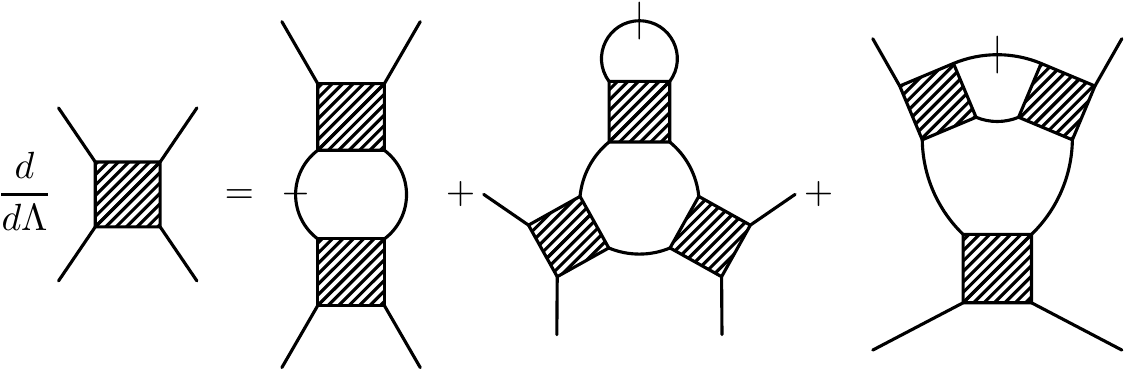}
	\caption{Simplified diagrammatic representation of the two-loop flow equation
for the two-particle vertex. The second and third diagrams on the right hand
side can be classified as two-loop contributions with non-overlapping and
overlapping loops, respectively.}
	\label{fig3:FlowEq2P_TwoLoop}
\end{figure}

The flow equation for the three-particle vertex can be written schematically
as\footnote{The flow equation for $\Gamma^{(6)\Lambda}$ including all indices
and contributions in $\mathcal O((\V{})^3)$ follows from
Eqs.~\eqref{eq:fRG:Gamma6-3P-1} and~\eqref{eq:fRG:Gamma6-3P-2} by applying the
derivative $\partial_{\Lambda,S}$ to the right hand sides.}
\begin{equation}
\begin{split}
	\frac{d}{d\Lambda}\Gamma^{(6)\Lambda} =& \V{} \partial_{\Lambda,S}(\G{} \G{})
\Gamma^{(6)\Lambda} +
\SL{} \Gamma^{(8)\Lambda}\\
	&+ \partial_{\Lambda,S} \tr (\G{} \V{} \G{} \V{} \G{} \V{})
\end{split}
\label{eq:RGEq_Gamma6}
\end{equation}
and is illustrated diagrammatically in Fig.~\ref{fig2:FlowEq3P}. A solution
for
$\Gamma^{(6)\Lambda}$ in $\mathcal O((\V{})^3)$ is obtained by neglecting
the contributions in the first line of Eq.~\eqref{eq:RGEq_Gamma6}, because
these are at least of fourth order in the effective interaction. At the same
level of approximation, the scale-derivative $\partial_{\Lambda,S}$ in the
second line of Eq.~\eqref{eq:RGEq_Gamma6} can be replaced by a full
scale-derivative $\frac{d}{d\Lambda}$ that acts also on the
self-energy and the vertex. The latter modification generates terms at least of 
$\mathcal O((\V{})^4)$. The resulting flow equation~\cite{Salmhofer2004}
\begin{equation}
	\frac{d}{d\Lambda}\Gamma^{(6)\Lambda} = \frac{d}{d\Lambda} \tr (\G{} \V{}
\G{} \V{} \G{} \V{})
\label{eq:RGEq_Gamma6O3}
\end{equation}
can straightforwardly be integrated because the right hand side is a total
derivative, yielding
\begin{equation}
	\Gamma^{(6)\Lambda} = \tr(\G{} \V{} \G{} \V{} \G{} \V{})
\label{eq:Gamma6O3schematic}
\end{equation}
in a system with only two-particle interactions at the microscopic level. In
more detail, the three-particle vertex in $\mathcal O((\V{})^3)$ reads
\begin{widetext}
\begin{gather}
	\Gamma^{(6)\Lambda}_{\alpha\beta a b \gamma\delta} =
(\Gamma^{(6)\Lambda}_{\alpha\beta a b \gamma\delta})_{(1)} +
(\Gamma^{(6)\Lambda}_{\alpha\beta a b \gamma\delta})_{(2)},\\
	\begin{split}
		(\Gamma^{(6)\Lambda}_{\alpha\beta a b \gamma\delta})_{(1)} =
-\hspace{-1ex}\sum_{c,d,e,f,m,n}\hspace{-1ex}& \G{cd} \G{ef} \G{mn}
\bigl(\V{\alpha n e
\delta} \V{\beta d m b} \V{a f c \gamma} - \V{\alpha n e \delta} \V{a d m b}
\V{\beta f c \gamma} + \V{\alpha n e \gamma} \V{a d m b} \V{\beta f c
\delta} -\V{\alpha n e \gamma} \V{\beta d m b} \V{a f c \delta}\\[-2ex]
	& \hspace{4em} + \V{\alpha n e \delta} \V{a d m \gamma} \V{\beta f c b}
-\V{\alpha n e \delta} \V{\beta d m \gamma} \V{a f c b} - \V{\alpha n e \gamma}
\V{a d m \delta} \V{\beta f c b} + \V{\alpha n e \gamma} \V{\beta d m \delta}
\V{a f c b} \\
	& \hspace{4em} - \V{\alpha n e b} \V{a d m \gamma} \V{\beta f c \delta} +
\V{\alpha n e b} \V{\beta d m \gamma} \V{a f c \delta} +
\V{\alpha n e b} \V{a d m \delta} \V{\beta f c \gamma} - \V{\alpha n e b}
\V{\beta d m \delta} \V{a f c \gamma}\bigr),
	\label{eq:fRG:Gamma6-3P-1}
	\end{split} \\
\begin{split}
	\hspace{-1ex}(\Gamma^{(6)\Lambda}_{\alpha\beta a b \gamma\delta})_{(2)} =
-\hspace{-1ex}\sum_{c,d,e,f,m,n}\hspace{-1ex}& \G{cd} \G{ef} \G{mn}
\bigl(\V{\beta a c m}
\V{f d \gamma \delta} \V{\alpha n e b} + \V{a \alpha c m} \V{f d \gamma \delta}
\V{\beta n e b} + \V{\alpha \beta c m} \V{f d \gamma \delta} \V{a n e b} +
\V{\alpha\beta c m} \V{f d b \gamma} \V{a n e \delta}\\[-2ex]
 &\hspace{4em} + \V{\alpha \beta c m} \V{f d \delta b} \V{a n e \gamma} +
\V{\beta a c m} \V{f d b \gamma} \V{\alpha n e \delta} + \V{\beta a c m}
\V{f d \delta b} \V{\alpha n e \gamma} + \V{a \alpha c m} \V{f d b \gamma}
\V{\beta n e \delta}\\
	& \hspace{4em} + \V{a \alpha c m} \V{f d \delta b} \V{\beta b e \gamma}\bigr).
	\label{eq:fRG:Gamma6-3P-2}
\end{split}
\end{gather}
\end{widetext}
In Eq.~\eqref{eq:fRG:Gamma6-3P-1} all internal propagators run into the same
direction, while this is not the case in Eq.~\eqref{eq:fRG:Gamma6-3P-2}.
Inserting these expressions in Eq.~\eqref{eq:fRG:dVdL} yields the flow equation
for the two-particle vertex at two-loop level, which is illustrated
schematically in Fig.~\ref{fig3:FlowEq2P_TwoLoop}.

The two-loop diagrams can be classified in contributions with non-overlapping
and overlapping loops, which have the topological structure of the second and
third diagram on the right hand side of Fig.~\ref{fig3:FlowEq2P_TwoLoop},
respectively. In two-loop diagrams with non-overlapping loops, the single-scale 
propagator in Eq.~\eqref{eq:fRG:dVdL} contracts two indices at the same vertex. 
As discussed by Katanin,\cite{Katanin2004a} these contributions can be 
rewritten as one-loop diagrams with an insertion of $\dot\Sigma^\Lambda$. 
Exploiting the relation $\dot G^\Lambda = \SL{} + \G{} \dot\Sigma^\Lambda \G{}$, 
they allow to replace the single-scale propagators in the one-loop contributions 
in Eq.~\eqref{eq:fRG:dVdL} by scale-differentiated full propagators.

The two-loop contributions with overlapping loops can be treated in a similar
fashion after decomposing the vertex in interaction channels, which is discussed
in the next section.

\subsection{Channel-decomposed flow equations at two-loop level:
\texorpdfstring{$\boldsymbol {\dot \Gamma}$}{Gamma-dot} scheme}
\label{subsec:TwoLoopChannelDecomp}
In this section, we discuss how the two-loop contributions with overlapping
loops can be reformulated effectively as one-loop contributions, and derive a 
channel-decomposition scheme for the vertex and the flow equations at two-loop 
level. The latter allows to extend the one-loop schemes by 
Karrasch~\etal~\cite{Karrasch2008} and Husemann and 
Salmhofer~\cite{Husemann2009} for the symmetric phase, by the author and 
Metzner for singlet superconductors,\cite{Eberlein2010} and by Maier and 
Honerkamp for antiferromagnets.\cite{Maier2012}

The basic idea is to rewrite the insertions of two vertices that are 
connected by a full and a single-scale propagator in the two-loop contributions 
with overlapping loops (see third diagram on the right hand side of 
Fig.~\ref{fig3:FlowEq2P_TwoLoop}) as one-loop contribution 
(see first diagram on right hand side of Fig.~\ref{fig3:FlowEq2P_TwoLoop}). In 
order to make use of this idea, the vertex has to be decomposed in interaction 
channels and the diagrams in the flow equation have to be assigned to channels 
according to their leading singular dependence on external momenta and 
frequencies. 

\begin{figure}
	\centering
	\includegraphics[width=0.9\linewidth]{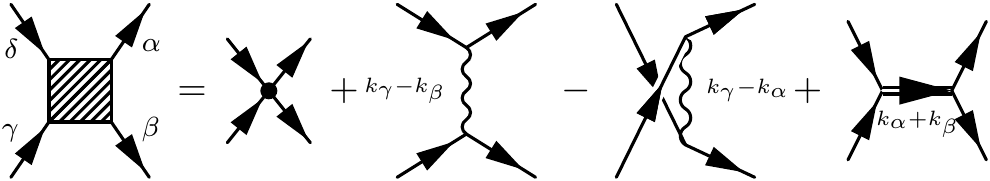}
	\caption{Diagrammatic representation of the decomposition of the (Nambu)
two-particle vertex in bare interaction, particle-hole channels and
particle-particle channel.}
	\label{fig4:ChannelDecomposition}
\end{figure}

For a decomposition in interaction channels, the two-particle vertex is written
as a sum of several terms, where each describes a possibly singular
dependence of the vertex on momenta and frequencies. Assuming translation 
invariance, the multiindex notation of section~\ref{subsec:TL_FlowEq} is 
specialized to
\begin{equation}
	\V{\alpha\beta\gamma\delta} \equiv \V{s_\alpha s_\beta s_\gamma
s_\delta}(k_\alpha, k_\beta, k_\gamma, k_\delta),
\end{equation}
where $\V{}$ is nonzero only for $k_\alpha + k_\beta = k_\gamma + k_\delta$.
For the sake of compactness of the presentation, we stick to the multiindex
notation in the major part of this section. The channel-decomposition of the
vertex reads
\begin{equation}
\begin{split}
	\V{\alpha \beta \gamma \delta}& = u_{\alpha\beta\gamma\delta} +
\VPH{\alpha\beta\gamma\delta} - \VPH{\beta \alpha \gamma \delta} +
\VPP{\alpha\beta\gamma\delta},
\end{split}
\label{eq:ChannelDecomposition}
\end{equation}
where $u$ is the microscopic interaction and 
\begin{gather}
	\VPH{\alpha\beta\gamma\delta} = \VPH{s_\alpha s_\beta s_\gamma
s_\delta}\bigl(k_\gamma - k_\beta; \tfrac{k_\alpha +
k_\delta}{2},\tfrac{k_\beta+k_\gamma}{2}\bigr),\\
	\VPP{\alpha\beta\gamma\delta} = \VPP{s_\alpha s_\beta s_\gamma
s_\delta}\bigl(k_\alpha + k_\beta; \tfrac{k_\alpha-k_\beta}{2},
\tfrac{k_\delta-k_\gamma}{2}\bigr)
\end{gather}
describe fluctuation corrections in the (Nambu) particle-hole and (Nambu)
particle-particle channel, respectively. The first argument of
$V^{\text{PH},\Lambda}$ and $V^{\text{PP},\Lambda}$ describes the possibly
singular dependence on the transfer momentum, while the last two momentum
arguments describe dependences on fermionic relative momenta. The decomposition
of the vertex  is illustrated diagrammatically in
Fig.~\ref{fig4:ChannelDecomposition}.

Flow equations for the effective interactions in the particle-hole and
particle-particle channels are derived by inserting the ansatz
Eq.~\eqref{eq:ChannelDecomposition} into the flow equation~\eqref{eq:fRG:dVdL}
and assigning diagrams to interaction channels according to their leading
singular dependence on external momenta. The one-loop contributions (\ie,
Eq.~\eqref{eq:fRG:dVdL} without the last term) are assigned to
interaction channels as in the channel-decomposition schemes at one-loop level
in such a way that the transfer momentum is transported through the diagrams by
the fermionic propagators. The first and second contributions in the square
bracket in Eq.~\eqref{eq:fRG:dVdL} are assigned to the direct and crossed
particle-hole channel, respectively. The third contribution is assigned to the
particle-particle channel. This yields
\begin{gather}
\begin{split}
	\bigl(\frac{d}{d\Lambda}\VPH{\alpha \beta \gamma
\delta}\bigr)_\text{1L} &= \sum_{a,b,c,d} \partial_{\Lambda,S}(\G{ab} \G{cd})
\V{\alpha b c \delta} \V{d \beta\gamma a} \\
	& \equiv \dVPHdL{\alpha \beta \gamma \delta},
\end{split}\label{eq:dVPH_dL_1L}\\
\begin{split}
	\bigl(\frac{d}{d\Lambda} \VPP{\alpha \beta \gamma \delta}\bigr)_\text{1L} &=
-\frac{1}{2} \sum_{a,b,c,d} \partial_{\Lambda,S}(\G{ab}\G{cd}) \V{\alpha\beta c
a} \V{b d \gamma \delta} \\
	& \equiv \dVPPdL{\alpha \beta \gamma \delta}
\end{split}\label{eq:dVPP_dL_1L}
\end{gather}
After rewriting the two-loop diagrams with non-overlapping loops as one-loop
diagrams with $\dot\Sigma$-insertions, they are assigned similarly according to
the transfer momentum in the fermionic loops,
\begin{gather}
	\bigl(\frac{d}{d\Lambda}\VPH{\alpha \beta \gamma
\delta}\bigr)_{\dot \Sigma} = \sum_{a,b,c,d} \partial_{\Lambda,\Sigma}(\G{ab}
\G{cd}) \V{\alpha b c \delta} \V{d \beta\gamma a},\\
	\bigl(\frac{d}{d\Lambda} \VPP{\alpha \beta \gamma \delta}\bigr)_{\dot \Sigma}
= -\frac{1}{2} \sum_{a,b,c,d} \partial_{\Lambda,\Sigma}(\G{ab}\G{cd})
\V{\alpha\beta c a} \V{b d \gamma \delta},
\end{gather}
where $\partial_{\Lambda,\Sigma}$ is a shorthand for a $\Lambda$-derivative
that acts on the fermionic self-energy only, $\partial_{\Lambda,\Sigma}
\G{} = \G{} \dot\Sigma^\Lambda \G{}$.

\begin{figure}
	\centering
	\includegraphics[width=\linewidth]{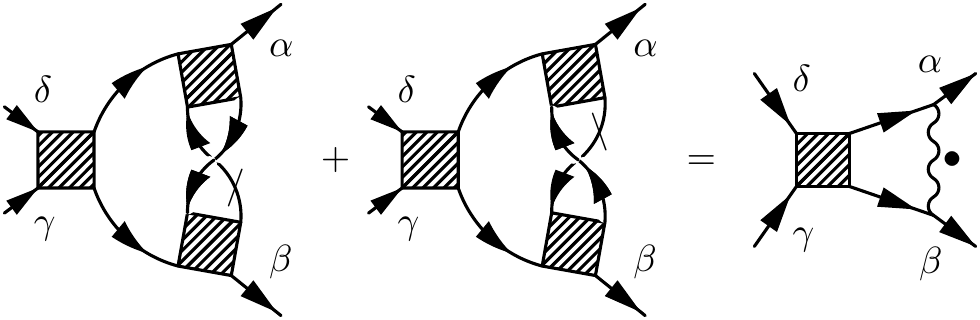}
	\caption{Illustration of the reorganization of two-loop diagrams with
overlapping loops in a one-loop diagram with a scale-differentiated effective
interaction.}
	\label{fig5:DiagrammCollectionExample}
\end{figure}
Before assigning the two-loop contributions with overlapping loops to the
interaction channels, it is convenient to rewrite them effectively as one-loop
diagrams. This is possible after expressing insertions of two vertices that are
connected by a full and a single-scale propagator as one-loop scale-derivatives
of effective interactions. Consider as an example the renormalization
contribution to the two-particle vertex arising from the first two terms in the
first line of Eq.~\eqref{eq:fRG:Gamma6-3P-2}. After insertion into
Eq.~\eqref{eq:fRG:dVdL}, these contributions read
\begin{equation}
	-\sum\SL{ba}\G{cd}\G{ef}\G{mn} (\V{\beta a c m} \V{\alpha n e b} + \V{a\alpha
c m} \V{\beta n e b}) \V{fd\gamma\delta}.
\end{equation}
Renaming summation indices and exploiting the antisymmetry of the vertex under
particle exchange, this expression can be rewritten as
\begin{gather*}
-\sum \V{f d \gamma \delta} \G{ef}\G{cd} \underbrace{(\SL{ba}\G{mn} + \G{ba}
\SL{mn}) \V{\alpha n b e} \V{a \beta c m}}_{=\bigl(\frac{d}{d\Lambda}\VPH{\alpha
\beta c e}\bigr)_\text{1L}}
\end{gather*}
\begin{gather}=-\sum_{c,d,e,f} \bigl(\frac{d}{d\Lambda} 
\VPH{\alpha \beta c
e}\bigr)_\text{1L} \G{ef} \G{cd} \V{fd \gamma \delta},
\end{gather}
where Eq.~\eqref{eq:dVPH_dL_1L} was exploited. This reorganization is
illustrated diagrammatically in Fig.~\ref{fig5:DiagrammCollectionExample} and 
works similarly for all contributions in Eq.~\eqref{eq:fRG:dVdL} after inserting
Eqs.~\eqref{eq:fRG:Gamma6-3P-1} and~\eqref{eq:fRG:Gamma6-3P-2}. Collecting terms
yields the two-loop flow equation for the vertex expressed effectively as a
one-loop equation,
\begin{widetext}
\begin{equation}
\begin{split}
	\frac{d}{d\Lambda} \V{\alpha\beta\gamma\delta} &= \sum_{a,b,c,d}
\Bigr[\frac{d}{d\Lambda}(\G{ab} \G{cd}) (\V{\alpha b c \delta} \V{d \beta\gamma
a} - \V{\beta b c \delta} \V{d\alpha\gamma a}) -
\frac{1}{2} \frac{d}{d\Lambda}(\G{ab}\G{cd}) \V{\alpha\beta c a} \V{b d
\gamma \delta}\Bigr]\\
	&+\sum_{a,b,c,d} \G{ab}\G{cd}\Bigl[ \V{\alpha b c \delta}
\dVPPdL{\beta d a \gamma} %
- \V{\alpha b c \gamma} \dVPPdL{\beta d a \delta} %
+ \dVPPdL{\alpha b c \delta} \V{\beta d a \gamma} %
- \dVPPdL{\alpha b c \gamma} \V{\beta d a \delta} %
 - \V{\alpha b c \delta} \dVPHdL{\beta d \gamma a} \\
&\hspace{5em}
+ \V{\alpha b c \gamma} \dVPHdL{\beta d \delta a} %
+ \dVPHdL{\alpha d \gamma a} \V{\beta b c \delta} %
- \dVPHdL{\alpha d \delta a} \V{\beta b c \gamma}
- \V{\alpha \beta c a} \dVPHdL{b d \gamma \delta} %
- \dVPHdL{\alpha \beta c a} \V{b d \gamma \delta}\Bigr].
\end{split}
\label{eq:fRG:dVdL_2L}
\end{equation}
\end{widetext}

\begin{figure}
	\centering
	\subfigure[][]{\includegraphics[scale=.7]{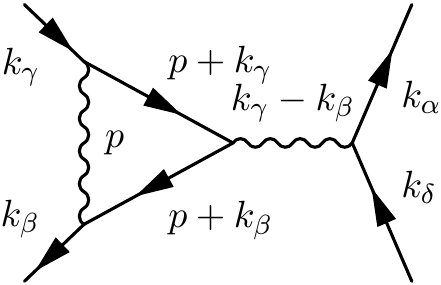}%
	\label{fig6a:VertexCorr}}\\
	\subfigure[][]{\includegraphics[scale=.7]{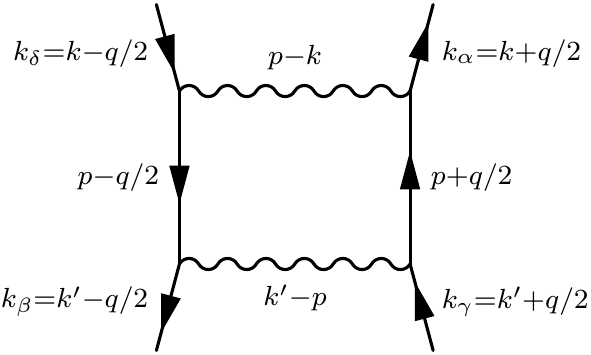}%
	\label{fig6b:Box}}
	\subfigure[][]{\includegraphics[scale=.7]{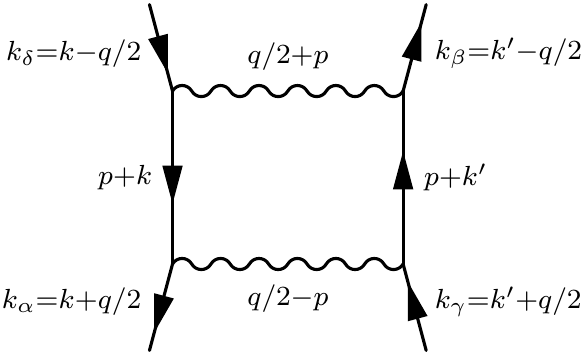}%
	\label{fig6c:TBE}}
	\caption{Examples for (a) vertex correction and (b,c) box diagrams. At 
two-loop level, the scale-derivative acts on the effective interactions that
transport the loop momentum $p$. The transfer momentum is transported through
the diagram by the fermionic propagators in (a) and (b), while it is transported
by the effective interactions in (c).}
	\label{fig6:DiagramExamples}
\end{figure}
In order to make use of this reorganization, we insert the decomposition of the
vertex in interaction channels Eq.~\eqref{eq:ChannelDecomposition} on both 
sides of Eq.~\eqref{eq:fRG:dVdL_2L} and assign diagrams to interaction channels
according to their leading singular dependence on external momenta. The
assignment of the contributions in the first line of Eq.~\eqref{eq:fRG:dVdL_2L}
was discussed above. After inserting the decomposition of the vertex, the second
and third lines contain diagrams that can be classified in two-loop vertex
correction diagrams (for an example see Fig.~\ref{fig6a:VertexCorr}) and 
two-loop box diagrams (for examples see Figs.~\ref{fig6b:Box} 
and~\subref{fig6c:TBE}). No two-loop propagator renormalization diagrams appear 
as a consequence of the topological structure of the two-loop diagrams with
overlapping loops.

\begin{figure}
	\centering
	\includegraphics[width=\linewidth]{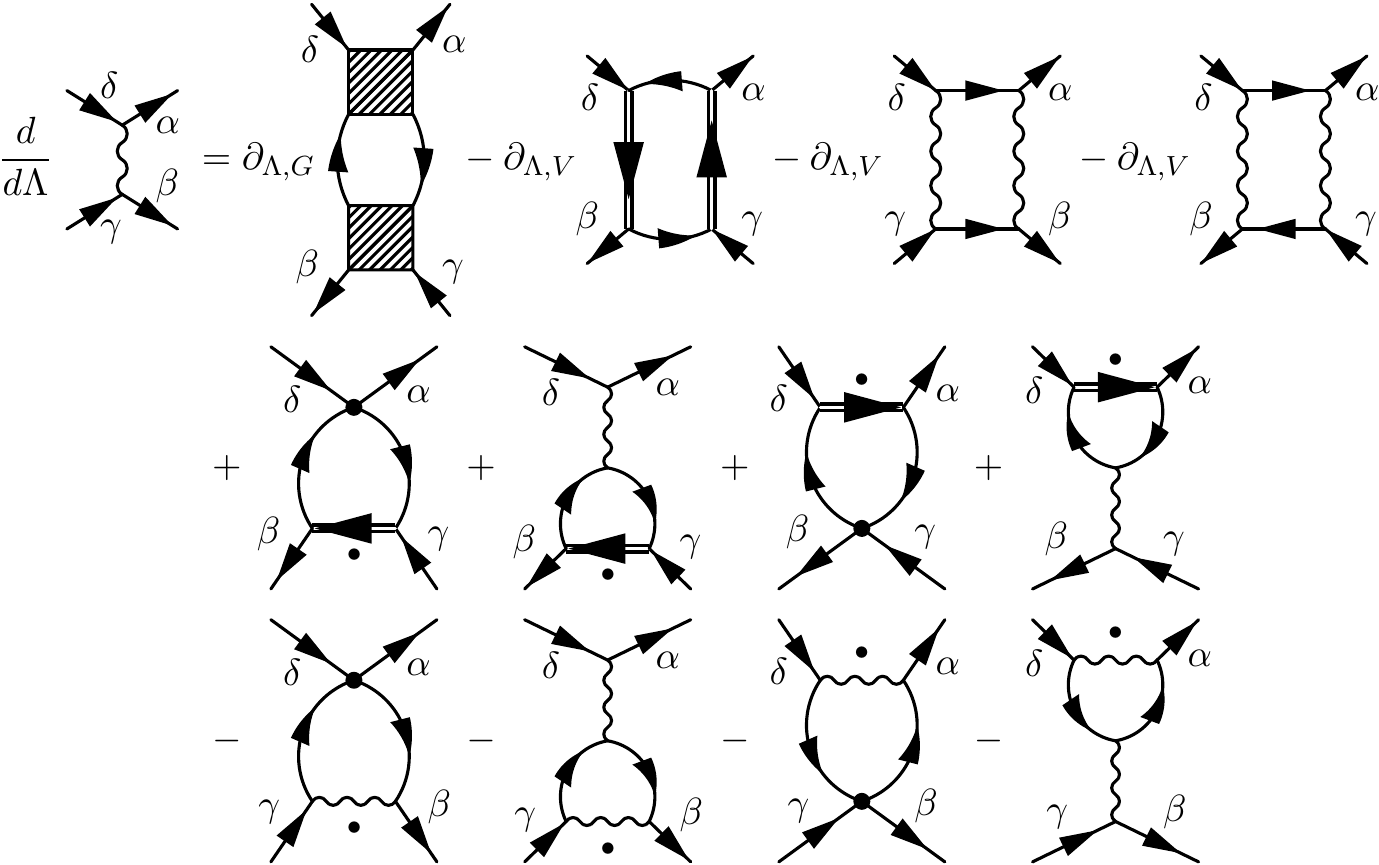}
	\caption{Diagrammatic representation of the two-loop renormalization group 
equation for the effective interaction in the (Nambu) particle-hole channel 
$\VPH{}$. The first term on the right hand side represents the one-loop
contributions where $\partial_{\Lambda,G} = \partial_{\Lambda,S} + 
\partial_{\Lambda,\Sigma}$ acts on fermionic propagators. The
other terms in the first line represent two-loop box diagrams where
$\partial_{\Lambda,V}$ is a shorthand for a scale-derivative acting on
effective interactions and yielding the one-loop result. The other terms
represent two-loop vertex correction diagrams where the effective interactions
with dots represent the one-loop contribution.}
\label{fig7:RGDE-Vertex-PH}
\end{figure}

\begin{figure}
	\centering
	\includegraphics[width=\linewidth]{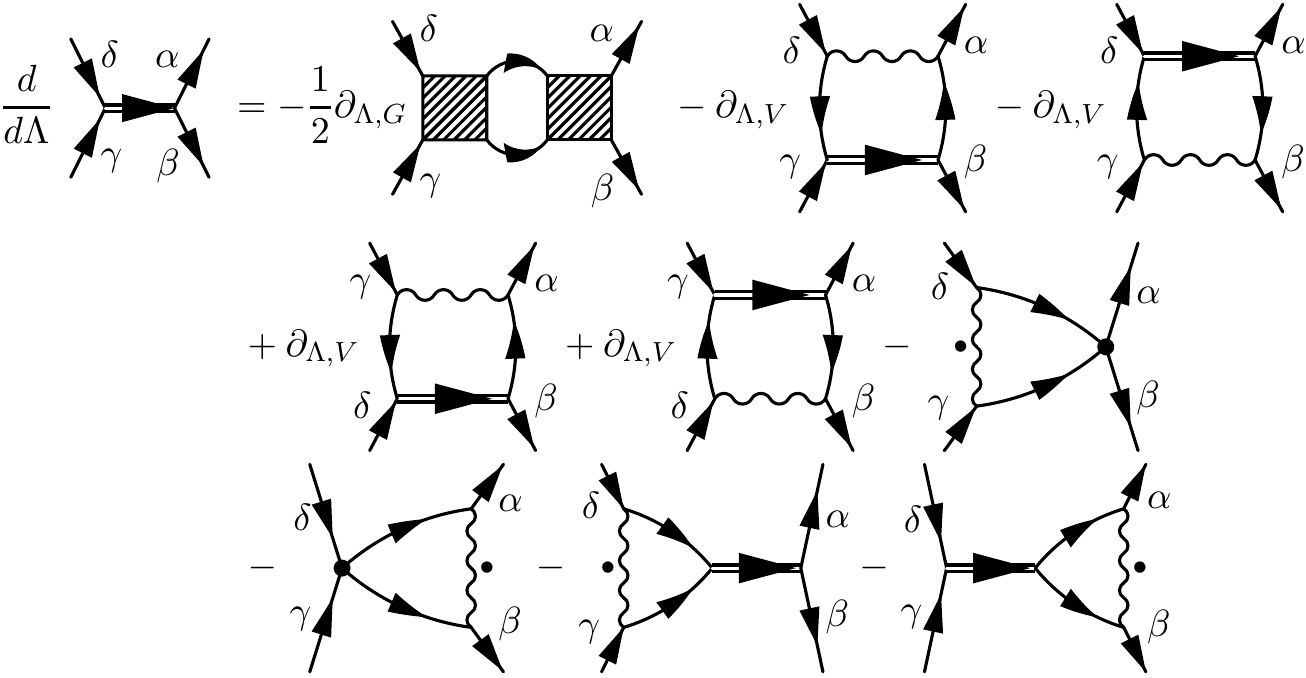}
	\caption{Diagrammatic representation of the two-loop renormalization group 
equation for the effective interaction in the (Nambu) particle-particle channel 
$\VPP{}$. The first term on the right hand side represents the one-loop
contribution. The last term in the second line and the contributions in the
third line represent two-loop vertex correction diagrams. The other
contributions are two-loop box diagrams. The notation is the same as in
Fig.~\ref{fig7:RGDE-Vertex-PH}.}
	\label{fig8:RGDE-Vertex-PP}
\end{figure}

Like at one-loop level, the vertex correction diagrams are assigned to
interaction channels according to the transfer momentum in the fermionic loop
and in one effective interaction. The singular dependence on momentum in the
other effective interaction is integrated and does not give rise to singular
renormalization contributions in the ground state of a fermionic $s$-wave
superfluid (see section~\ref{subsec:IREst}). Differently from one-loop level, 
the two-loop box diagrams are assigned to interaction channels in such a
way that the transfer momentum is transported through the diagram by the
effective interactions (as shown in the example in Fig.~\ref{fig6c:TBE}). The
reason is that these diagrams become important close to and below the critical
scale,\cite{Salmhofer2001} where the effective interactions and their
scale-derivatives already developed a strong
dependence on momentum and frequency. The assignment according to the
``bosonic'' singularity yields a better treatment of the strong momentum and
frequency dependence of the scale-differentiated effective interactions. In
section~\ref{subsec:IREst}, we demonstrate that this assignment allows to
capture the singular renormalization of the amplitude mode by long-range phase
fluctuations in a fermionic $s$-wave superfluid. The two-loop renormalization 
contributions to the effective interaction in the (Nambu) particle-hole and 
particle-particle channels read
\begin{widetext}
	\begin{gather}
\begin{split}
		\bigl(\frac{d}{d\Lambda}\VPH{\alpha \beta \gamma \delta}\bigr)_\text{2L} =
\sum_{a,b,c,d} \G{ab} \G{cd} &\Bigl[\bigl(u_{\alpha b c \delta} + \VPH{\alpha b
c \delta}\bigr) \bigl(\dVPPdL{d \beta \gamma a} -
\dVPHdL{\beta d\gamma a}\bigr) +
\bigl(\dVPPdL{\alpha b c \delta} - \dVPHdL{b \alpha c
\delta}\bigr) \bigl(u_{d \beta \gamma a} + \VPH{d \beta \gamma
a}\bigr)\\
& - \frac{d}{d\Lambda}\bigl(\VPP{\alpha b c \gamma} \VPP{d \beta \delta a}
+ \VPH{\alpha \beta c a} \VPH{b d \gamma \delta} + \VPH{\alpha d \gamma a}
\VPH{b \beta c \delta}\bigr)_\text{1L}\Bigr],
\end{split}\\
\begin{split}
	\bigl(\frac{d}{d\Lambda}\VPP{\alpha\beta\gamma\delta}\bigr)_\text{2L} =
-\sum_{a,b,c,d} \G{ab}\G{cd} &\Bigl[\bigl(u_{\alpha\beta c a} + \VPP{\alpha
\beta c a}\bigr) \dVPHdL{b d \gamma \delta} + \dVPHdL{\alpha\beta c
a} \bigl(u_{b d \gamma \delta} + \VPP{b d \gamma \delta}\bigr)\\
	&+ \frac{d}{d\Lambda}\bigl(\VPH{\alpha b \delta c} \VPP{\beta d a \gamma} +
\VPP{\alpha d a \delta} \VPH{\beta b \gamma c} - \VPH{\alpha b \gamma c}
\VPP{\beta d a \delta} - \VPP{\alpha d a \gamma} \VPH{\beta b \delta
c}\bigr)_\text{1L}\Bigr].
\end{split}
	\end{gather}
\end{widetext}
Taking into account the one-loop contributions and the two-loop contributions
with non-overlapping loops yields the channel-decomposed flow equations at 
two-loop level
\begin{gather}
\begin{split}
	\frac{d}{d\Lambda}\VPH{\alpha \beta \gamma
\delta} &=
\bigl(\frac{d}{d\Lambda}\VPH{\alpha \beta \gamma \delta}\bigr)_\text{1L} +
\bigl(\frac{d}{d\Lambda}\VPH{\alpha \beta \gamma \delta}\bigr)_{\dot \Sigma}\\
	& + \bigl(\frac{d}{d\Lambda}\VPH{\alpha \beta \gamma 
\delta}\bigr)_\text{2L},
\end{split}\\
\begin{split}
\frac{d}{d\Lambda}\VPP{\alpha \beta \gamma \delta} &=
\bigl(\frac{d}{d\Lambda}\VPP{\alpha \beta \gamma \delta}\bigr)_\text{1L} +
\bigl(\frac{d}{d\Lambda}\VPP{\alpha \beta \gamma \delta}\bigr)_{\dot \Sigma} \\
 & + \bigl(\frac{d}{d\Lambda}\VPP{\alpha \beta \gamma \delta}\bigr)_\text{2L}.
\end{split}
\end{gather}
These flow equations are illustrated diagrammatically in
Figs.~\ref{fig7:RGDE-Vertex-PH} and~\ref{fig8:RGDE-Vertex-PP}. The two-loop
contributions are computed using $\bigl(\frac{d}{d\Lambda}\VPH{\alpha \beta
\gamma \delta}\bigr)_\text{1L}$ and $\bigl(\frac{d}{d\Lambda}\VPP{\alpha \beta
\gamma \delta}\bigr)_\text{1L}$.

In the next section, we give a few remarks about the assignment of diagrams
in the two-loop channel-decomposition scheme. 

\subsection{Discussion of assignment of diagrams}
The channel-decomposition scheme derived in 
section~\ref{subsec:TwoLoopChannelDecomp} is exact to the third order in the
effective interaction. Due to the use of an approximation for the three-particle
vertex of that order for its derivation, no $\dot\Sigma$-insertions appear in
the scale-differentiated effective interactions in the two-loop contributions.
However, within the same order of approximation the $\dot\Sigma$-insertions can
be added to the scale-differentiated effective interactions in the two-loop
contributions, because this introduces terms of $\mathcal O((\V{})^4)$. The
resulting channel-decomposition scheme would correspond to the two-loop flow
equations for the two-particle vertex proposed by Veschgini and
Salmhofer.\cite{Veschgini2013} For the attractive Hubbard model, we checked
such an assignment and found only minor differences in the results in the
presence of a not too small external pairing field. Note however that this may
change in other contexts. For pairing field flows, including the
$\dot\Sigma$-insertions in the two-loop contributions leads to a somewhat
stronger renormalization of exchange propagators, but qualitatively similar
results. In the numerical study of the attractive Hubbard model, we therefore
did not include the $\dot\Sigma$-insertions in the two-loop contributions.

The scheme presented in the last section is relatively compact, because we
only distinguished between singularities arising from fermionic propagators or
effective interactions when assigning diagrams to interaction channels. For very
small external pairing fields, it may be advantageous to further distinguish
between one-loop box diagrams with normal or anomalous fermionic propagators.
The reason is that the anomalous self-energy receives strong renormalizations
in the pairing field flow due to the singular behavior of the amplitude mode.
Insertions of the resulting $\dot\Sigma$ in the one-loop box diagrams may lead
to artificial logarithmic singularities in non-Cooper channels when assigned as
described above. These can be avoided by assigning the one-loop box diagrams
with anomalous fermionic propagators in such a way that the transfer momentum is
transported through the diagrams by the effective interactions. The one-loop box
diagrams with normal fermionic propagators do not cause difficulties and should
be assigned as described above. These subtleties matter only for very small
external pairing fields (well beyond those that were accessible in the
numerics). Therefore we decided to present the two-loop channel-decomposition
scheme in the simpler form as above, but consider the more sophisticated
version in the estimates in section~\ref{subsec:IREst}. We checked that the
alternative assignment of one-loop box diagrams with anomalous propagators
indeed yields very similar results in the numerically accessible range of
external pairing fields.

\section{Attractive Hubbard model}
\label{sec:AHM}
In this section, we study the ground state of the attractive Hubbard model on
the square lattice as a prototype for a singlet superfluid~\cite{*[{For a
review, see }] [{}] Micnas1990} using the two-loop channel-decomposition
scheme. Progress in experiments with cold atoms in optical lattices sparked
renewed research interest in this model in the last decade, because it can be
simulated in such systems.\cite{Bloch2008,Esslinger2010}

The attractive Hubbard model describes spin-$\frac{1}{2}$ fermions with a local 
attractive interaction on a lattice. Its Hamiltonian is
\begin{equation}
	H = \sum_{i,j,\sigma} t_{ij} c^\dagger_{i\sigma} c_{j\sigma} + U \sum_i
n_{i\uparrow} n_{i\downarrow},
\end{equation}
where $c^\dagger_{i\sigma}$ and $c_{i\sigma}$ are creation and annihilation
operators for fermions with spin orientation $\sigma$ on
lattice site $i$. The interaction parameter $U$ is negative. The hopping of
fermions is restricted to nearest- and next-nearest-neighbour sites with
amplitudes $-t$ and $-t'$, respectively, yielding the dispersion relation
\begin{equation}
\epsilon(\boldsymbol k) = -2 t (\cos k_x + \cos k_y) -4t' \cos k_x \cos 
k_y.\\[0.5ex]
\end{equation}
In the following, we use $t\equiv 1$ as the unit of energy.

The ground state of this model is an $s$-wave spin-singlet superfluid at any
fermionic density. For $t' = 0$ and $n = 1$ (half-filling), superfluidity is
degenerate with charge-density wave order. The model has been studied in its
ground state and at finite temperatures with a variety of methods, including
resummations of perturbation
theory,\cite{Fresard1992,Pedersen1997,Letz1998,Keller1999}
quantum~\cite{Randeria1992,Trivedi1995,dosSantos1994,Singer1996} and
variational~\cite{Tamura2012} Monte Carlo methods, dynamical mean-field
theory,\cite{Keller2001,Keller2002,Capone2002} and functional renormalization
group.\cite{Gersch2008,Strack2008,Eberlein2013,Obert2013}

\subsection{Parametrization and approximations}
\label{subsec:Approx}
We now describe the approximate parametrization of the interaction vertex and
self-energy, which allow for a numerical solution of the two-loop flow
equations. The parametrizations are similar to those of
Ref.~\onlinecite{Eberlein2013}.

In order to describe the superfluid state, we use Nambu fields (see
Eq.~\eqref{eq:NambuRepresentation}). In this representation, the fermionic
propagator is a $2\times 2$ matrix and reads
\begin{equation}
	\boldsymbol G^\Lambda(k) = \begin{pmatrix}
			G^\Lambda_{++}(k)	&	G^\Lambda_{+-}(k)\\
			G^\Lambda_{-+}(k)	&	G^\Lambda_{--}(k)
	                           \end{pmatrix}
=\begin{pmatrix}
			G^\Lambda(k)	&	F^\Lambda(k)\\
			F^{\Lambda\ast}(k)	&	-G^\Lambda(-k),
	                           \end{pmatrix}
\end{equation}
where $G^\Lambda(k)$ and $F^\Lambda(k)$ are its normal and anomalous components,
respectively. The propagator is connected with the regularized bare propagator
$\boldsymbol G_0^\Lambda$ and the Nambu self-energy $\boldsymbol \Sigma^\Lambda$
by the Dyson equation $(\boldsymbol G^\Lambda)^{-1} = (\boldsymbol
G_0^\Lambda)^{-1} - \boldsymbol \Sigma^\Lambda$. $\boldsymbol G_0^\Lambda$ reads
\begin{widetext}
\begin{equation}
	\bigl(\boldsymbol G^\Lambda_0(k)\bigr)^{-1} = \begin{pmatrix}
	i k_0 - \xi(\boldsymbol k) - \delta\xi^\Lambda(\boldsymbol k) +
R^\Lambda(k_0)	&	\Delta_0\\
	\Delta_0		&	i k_0 + \xi(\boldsymbol k) + \delta\xi^\Lambda(\boldsymbol k) +
R^\Lambda(k_0)
	\end{pmatrix},
\label{eq:NambuG0}
\end{equation}
\end{widetext}
where $\xi(\boldsymbol k) = \epsilon(\boldsymbol k) -\mu$ and
$\delta\xi^\Lambda(\boldsymbol k)$ is a counterterm. The regulator
function
\begin{equation}
	R^\Lambda(k_0) = i \operatorname{sgn}(k_0) \sqrt{k_0^2 + \Lambda^2} - i k_0
	\label{eq:Regulator}
\end{equation}
regularizes the fermionic singularities by replacing small frequencies $k_0$
with $|k_0| \ll \Lambda$ by $\operatorname{sgn} (k_0) \Lambda$. The external
pairing field $\Delta_0$ appears in the \emph{bare} propagator, because it
serves as a regulator for pairing field flows,\cite{Eberlein2013} in which
$\Delta_0$ is eliminated after integrating out the fermionic modes at $\Lambda >
0$. The Nambu self-energy is given by
\begin{equation}
	\boldsymbol \Sigma^\Lambda(k) = \begin{pmatrix}
	\Sigma^\Lambda(k)	&	\Delta_0 - \Delta^\Lambda(k)\\
	\Delta_0 - \Delta^{\Lambda\ast}(k)		&		-\Sigma^\Lambda(-k)
	                                \end{pmatrix},
\label{eq:NambuSelfenergy}
\end{equation}
where the off-diagonal entries are chosen in such a way that the fermionic gap
is $\Delta^\Lambda(k)$. The counterterm is related to the normal
component of the self-energy via
\begin{equation}
	\delta\xi^\Lambda(\boldsymbol k) + \Sigma^\Lambda(0,\boldsymbol k) = 0
\end{equation}
for $\boldsymbol k$ on the Fermi surface at all $\Lambda$, such that the Fermi
surface remains fixed during the flow.
Similar to Ref.~\onlinecite{Eberlein2013}, we neglect the momentum dependence of
the self-energy but keep its dependence on frequency,
\begin{align}
	\Delta^\Lambda(k) &= \Delta^\Lambda(k_0),		&		\Sigma^\Lambda(k) &=
\Sigma^\Lambda(k_0),		&		\delta\xi^\Lambda(\boldsymbol k) &=
\delta\xi^\Lambda.
\end{align}
The momentum dependence of the self-energy is not expected to be important at
low fermionic densities, where the Fermi surface is almost circular.
Furthermore, it lead only to minor changes of fRG flows for the weakly-coupled
repulsive Hubbard model at van Hove filling.\cite{Giering2012} By choosing the
external pairing field $\Delta_0$ to be real, we fix the phase of the anomalous
self-energy so that $\Delta^\Lambda$ is also real. The frequency dependence of
the self-energy is discretized on a grid of 30 points that is denser near $k_0 =
0$ and becomes sparser towards higher frequencies, with a maximal frequency
around 300. Cubic spline interpolation is used to determine the self-energy at
intermediate frequencies.

The interaction vertex is fully described by several coupling
functions:\cite{Eberlein2010,Eberlein2013} $C^\Lambda_{kk'}(q)$ and
$M^\Lambda_{kk'}(q)$ describe charge and spin fluctuations, respectively. The
amplitude and phase mode of the superfluid gap are described by
$A^\Lambda_{kk'}(q)$ and $\Phi^\Lambda_{kk'}(q)$. The imaginary part of the
normal interaction in the Cooper channel is denoted as $P''^\Lambda_{kk'}(q)$.
The real and imaginary part of the anomalous $(3+1)$ effective interaction is
described by $X'^\Lambda_{kk'}(q)$ and $X''^\Lambda_{kk'}(q)$, respectively. The
coupling functions are expanded in exchange propagators that describe the
singular dependence on the transfer momentum $q$ and fermion-boson vertices for
the more regular dependences on the fermionic relative momenta $k$ and $k'$. As
in Ref.~\onlinecite{Eberlein2013}, we restrict this expansion to the $s$-wave
channel and approximate the coupling functions by the following ansatz:
\begin{equation}
	\begin{split}
		C^\Lambda_{kk'}(q) &= C^\Lambda(q),\\
		M^\Lambda_{kk'}(q) &= M^\Lambda(q),\\
		A^\Lambda_{kk'}(q) &= A^\Lambda(q) g_a^\Lambda(k_0) g_a^\Lambda(k_0'),\\
		\Phi^\Lambda_{kk'}(q) &= \Phi^\Lambda(q) g_\phi^\Lambda(k_0)
g_\phi^\Lambda(k_0'),\\
		P''^\Lambda_{kk'}(q) &= P''^\Lambda(q),\\
		X'^\Lambda_{kk'}(q) &= X'^\Lambda(q),\\
		X''^\Lambda_{kk'}(q) &= X''^\Lambda(q).
	\end{split}
\end{equation}
In comparison to Ref.~\onlinecite{Eberlein2013}, we neglect the renormalization
of the fermion-boson vertices in the particle-hole channel,
for the anomalous $(3+1)$ effective interactions and for the imaginary part of
the normal interaction in the Cooper channel. $g_a^\Lambda(k_0)$ and
$g_\phi^\Lambda(k_0)$ are kept in order to obtain a meaningful frequency
dependence of the gap function and to improve the fulfillment of the Ward
identity for the global $U(1)$ charge symmetry. These simplifications are
justified because the neglected frequency-dependences of fermion-boson vertices
had only a minor influence on the flow at one-loop level. The effective
interactions in the Nambu particle-hole and Nambu particle-particle channel
then read
\begin{align}
	\VPH{s_1 s_2 s_3 s_4}(q; k,k') &= C^\Lambda(q) \mtau{3}{s_1 s_4} \mtau{3}{s_2
s_3} + M^\Lambda(q) \mtau{0}{s_1 s_4} \mtau{0}{s_2 s_3}\nonumber\\
	&+\frac{1}{2} A^\Lambda(q)g_a^\Lambda(k_0) g_a^\Lambda(k_0') \mtau{1}{s_1
s_4}\mtau{1}{s_2 s_3}\nonumber\\
	&+\frac{1}{2} \Phi^\Lambda(q)g_\phi^\Lambda(k_0) g_\phi^\Lambda(k_0')
\mtau{2}{s_1 s_4}\mtau{2}{s_2 s_3} \label{eq:VPHansatz}\\
	&+\frac{1}{2} P''^\Lambda(q) (\mtau{1}{s_1 s_4} \mtau{2}{s_2 s_3} -
\mtau{2}{s_1 s_4} \mtau{1}{s_2 s_3})\nonumber\\
	&+ X'(q) (\mtau{3}{s_1 s_4}\mtau{1}{s_2 s_3} + \mtau{1}{s_1 s_4} \mtau{3}{s_2
s_3})\nonumber\\
	&+ X''(q) (\mtau{3}{s_1 s_4}\mtau{2}{s_2 s_3} - \mtau{2}{s_1 s_4} \mtau{3}{s_2
s_3}),\nonumber\\
	\VPP{s_1 s_2 s_3 s_4}(q; k,k') &= 2 M^\Lambda(q) \mtau{2}{s_1 s_2}\mtau{2}{s_3
s_4} \label{eq:VPPansatz}
\end{align}
where $\mtau{i}{}$ are Pauli matrices ($i = 1,2,3$) and the unit matrix ($i =
0$). All functions in Eqs.~\eqref{eq:VPHansatz} and~\eqref{eq:VPPansatz} are
even functions of momentum. $P''^\Lambda$ and $X''^\Lambda$ are odd functions of
frequency, while all other functions are even. The frequency dependence of the
fermion-boson vertices is discretized like for the self-energy. For the
exchange propagators, we discretize the dependence on momenta and frequencies
on a three-dimensional grid and trilinear interpolation is used at
intermediate momenta and frequencies. The frequency dependence is discretized
with 40 frequencies between $q_0 = 0$ and 300, with grid points denser
at small frequencies. The momentum dependence is discretized with cylindrical
coordinates around $\boldsymbol q = \boldsymbol 0$ and $\boldsymbol \pi$,
similar to Ref.~\onlinecite{Husemann2009}. The angular dependences are resolved
with three angles between $0$ and $\pi/4$. At quarter-filling, the radial
dependence of the singular exchange propagators $A(q)$, $\Phi(q)$, $P''(q)$ and
$X''(q)$ around $\boldsymbol q = \boldsymbol 0$ is discretized with 25 points
between radius 0 and $\pi$, with denser distribution of points near
$|\boldsymbol q| = 0$. All other exchange propagators have a weaker dependence
on momentum and are thus described with only 10 points in the radial direction.

The flow of the exchange propagators and fermion-boson vertices
is extracted from the flow equations for the coupling functions as described in
Ref.~\onlinecite{Eberlein2013} by averaging the external fermionic momenta
$\boldsymbol k$ and $\boldsymbol k'$ over the Fermi surface for suitable
choices of the transfer momentum and the fermionic frequencies. The flow of the
exchange propagators is evaluated for $k_0 = k_0' = 0$. The renormalization
contributions to $g^\Lambda_a$ and $g^\Lambda_\phi$ are obtained after setting
$q = 0$ and $k_0' = 0$. The flow of the fermionic self-energy is evaluated
similarly by averaging the external fermionic momentum over the Fermi surface.

By discretizing the dependences on momenta and frequencies, the functional flow
equations were transformed into a system of around 20000 non-linear ordinary
differential equations with three-dimensional loop integrals on the right-hand
sides. These loop integrals were performed with an adaptive integration
algorithm. The system of differential equations was integrated using an
adaptive third-order Runge-Kutta routine. Depending on the parameters, the
numerical integration of a flow on 32 CPU cores required around three days at
one-loop level and between two and four weeks at two-loop level. Due to the
large number of flowing couplings and the fact that the computations of their
renormalization contributions are independent at a given scale, the flow
equations are well suited for parallelization.

The numerical integration of the flow equations was started at a large
finite scale $\Lambda_0 \approx 100$, which is of the order of several times the
band width. The fermionic modes for $\Lambda > \Lambda_0$ were treated in
second-order perturbation theory, yielding exchange propagators of
the order of $-U^2 / \Lambda_0$ for $q = 0$, so that the contributions from
$\Lambda > \Lambda_0$ are small compared to $U$. Treating the high energy scales
in perturbation theory also provides a well defined starting point for the flow
of the fermion-boson vertices, which were set to one at $\Lambda_0$. 
The normal self-energy receives a sizeable contribution from the tadpole diagram
at any finite $\Lambda_0$, yielding $\Sigma^{\Lambda_0} = -\delta\xi^{\Lambda_0}
\approx U/2 + \mathcal O(\Lambda_0^{-1})$. The anomalous self-energy
$\Delta^{\Lambda_0}$ is determined self-consistently from the gap equation at
scale $\Lambda_0$, but the corrections to $\Delta_0$ are small (of order
$\mathcal O(U / \Lambda_0)$).

Due to the truncation of the hierarchy of flow equations and the approximations
for the coupling functions, the Ward identity for the global $U(1)$ charge
symmetry is violated in the two-loop flows.\cite{Eberlein2013-thesis} We did
not systematically study how large these violation are quantitatively. For
computing the results in section~\ref{subsec:NumResults}, the Ward identity was
enforced by a projection of the coupling constants as in
Ref.~\onlinecite{Eberlein2013}. At low scales, this procedure effectively
amounts to determining the Goldstone mass from the Ward identity instead of the
flow equation.

\subsection{Analytical estimates for infrared behavior}
\label{subsec:IREst}
Before presenting results from numerical solutions of the flow equations, we
discuss the infrared behavior of the vertex in a fermionic $s$-wave superfluid
in the BCS regime at zero temperature. We assume that the fermionic modes at
$\Lambda > 0$ have been integrated out in the presence of an external pairing
field $\Delta_0$. The latter regularizes the phase mode of the superfluid gap
and is treated in a pairing field flow~\cite{Eberlein2013} in this section.
For simplicity, we assume a circular Fermi surface, but its shape is not
expected to influence the conclusions. The infrared  behavior at one-loop level
was discussed in Ref.~\onlinecite{Eberlein2013}. In this section, we discuss
only estimates for the most singular contributions to the flow at two-loop
level. The flow equations including all terms are rather lengthy and can be
found in Ref.~\onlinecite{Eberlein2013-thesis}.

At small transfer momenta and frequencies, the exchange propagators in the
Cooper channel and for the imaginary part of the anomalous (3+1) effective
interaction are well described by
\begin{align}\label{eq:ExPropAnsatz}
\begin{aligned}
	\Phi^{\Delta_0}(q) &\sim -\frac{1}{\Delta_0 + Z_\Phi^{\Delta_0} q_0^2 +
A_\Phi^{\Delta_0} \boldsymbol q^2},\\
	P''^{\Delta_0}(q) &\sim -\frac{q_0}{\Delta_0 + Z_{P''}^{\Delta_0} q_0^2 +
A_{P''}^{\Delta_0} \boldsymbol q^2},\\
	X''^{\Delta_0}(q) &\sim \frac{q_0}{\Delta_0 + Z_{X''}^{\Delta_0} q_0^2 +
A_{X''}^{\Delta_0} \boldsymbol q^2},\\
	A^{\Delta_0}(q) &\sim -\frac{1}{\sqrt{\Delta_0 + Z_A^{\Delta_0} q_0^2 +
A_A^{\Delta_0} \boldsymbol q^2}},
\end{aligned}
\end{align}
where the superscripts indicate that $\Delta_0$ is the flow parameter. The
ansätze for $\Phi^{\Delta_0}$, $P''^{\Delta_0}$ and $X''^{\Delta_0}$ are
consistent with a resummation of all chains of Nambu particle-hole
diagrams~\cite{Eberlein2013} and can be justified non-perturbatively for
$\Phi^{\Delta_0}$ and $P''^{\Delta_0}$ using Ward
identities.\cite{Castellani1997a,Pistolesi2004} The ansatz for $A^{\Delta_0}$
is consistent with the expected singular infrared behavior of the amplitude
mode in an interacting Bose gas and reproduces the singular infrared scaling
that was described in Refs.~\onlinecite{Strack2008,Obert2013} in terms of
divergent wave function renormalization factors. The above-mentioned works
indicate that the coefficients $A_i^{\Delta_0}$ and $Z_i^{\Delta_0}$ in the
ansätze remain finite when defined as above. We assume that all other exchange
propagators are less singular in the limit where the external pairing field
vanishes. Below we show that these assumptions are justified and the ansätze
consistent with the infrared behavior of the flow. The fermion-boson vertices
are set to one in this section, as they are not expected to influence the
singular behavior. 

In the presence of a superfluid gap and close to the Fermi surface, the
fermionic propagator behaves like
\begin{gather}
\begin{aligned}
	F^{\Delta_0}(k+p) &\approx \frac{1}{\Delta^{\Delta_0}},\\
	G^{\Delta_0}(k+p) &\approx -\frac{i p_0 + v_F \boldsymbol p \cdot
\boldsymbol e_{\boldsymbol k_F}}{(\Delta^{\Delta_0})^2}
\end{aligned}
\end{gather}
for small $p$ and $k = (0, \boldsymbol k_F)$, together with appropriate
ultraviolet cutoffs, where $v_F$ is the Fermi velocity and
$\boldsymbol e_{\boldsymbol k_F}$ a unit vector pointing in the direction of
$\boldsymbol k_F$. In this section, we neglect the normal self-energy and
assume that it can be subsumed into Fermi liquid like renormalization factors,
which do not influence the singular infrared behavior as they remain finite (see
below).

The most singular contributions at two-loop level arise from diagrams
involving scale-derivatives of the phase mode. The two-loop vertex
correction diagrams of this kind involve an integral of the form
\begin{equation}
	\intdrei{p} \partial_{\Delta_0} \Phi^{\Delta_0}(p),
\end{equation}
where we suppressed the fermionic propagators and the second effective
interaction. The integrals should be evaluated with some ultraviolet cutoff
arising from the decay of the fermionic propagators at high frequencies and the
lattice. In case the integrals do not cause problems at the upper integration
limit, we send the cutoffs to infinity as the interesting behavior arises from
the region around $p = 0$. Neglecting the contributions from the (finite)
$\Delta_0$-derivatives of the renormalization factors for the dependence on
momenta and frequencies, this integral yields
\begin{equation}
	\intdrei{p} \frac{1}{(\Delta_0 + Z^{\Delta_0}_\Phi p_0^2 + A^{\Delta_0}
p^2)^2} \sim \frac{1}{A^{\Delta_0}_\Phi \sqrt{Z^{\Delta_0}_\Phi \Delta_0}}.
\end{equation}
In the following, the renormalization factors for the momentum and frequency
dependences are suppressed, as they are assumed to be finite and non-singular in
the limit of a vanishing external pairing field and thus do not change the
infrared behavior qualitatively. When multiplying this contribution with a
finite effective interaction, as in the flow equations for the exchange
propagators in the particle-hole channel, it may give rise to non-analytic
behavior but not to divergences as a function of $\Delta_0$ after integrating
the flow. In the flow equations for the exchange propagators in the
particle-particle channel, more singular contributions appear either from
propagator renormalization or two-loop box diagrams.

The two-loop box diagrams are potentially more singular than the vertex
correction diagrams, as they contain loops with two exchange propagators for the
phase mode. The biggest change in the infrared behavior in comparison to the
one-loop approximation is found for the amplitude mode $A^{\Delta_0}$.
Evaluating the two-loop box diagram for external momenta $k = k' = (0,
\boldsymbol k_F)$, the leading contributions read
\begin{equation}
\begin{split}
	\frac{d}{d\Delta_0}&A^{\Delta_0}(0)|_\text{2L} \sim -\intdrei{p} \Bigl[
F(k+p)^2 \\ &\times \partial_{\Delta_0} \bigl(A^{\Delta_0}(p)^2 +
\Phi^{\Delta_0}(p)^2 - 8 X''^{\Delta_0}(p)^2 + P''^{\Delta_0}(p)^2\bigr)\\
&+8 \operatorname{Im} G(k+p) F(k+p) \partial_{\Delta_0}\bigl(X''^{\Delta_0}(p)
\Phi^{\Delta_0}(p)\bigr)\Bigr]\\
&+\ldots.
\end{split}
\label{eq:dA_dDelta0_TwoLoopLeading}
\end{equation}
The ellipsis represents less singular terms in $\Delta_0$ that either involve
less singular exchange propagators or where their singularities are suppressed
by momentum and frequency factors stemming from the fermionic propagators. The
most singular contribution arises from the squared exchange propagator for the
phase mode, for which an estimate yields
\begin{equation}
\begin{split}
	\frac{d}{d\Delta_0}A^{\Delta_0}(0)|_\text{2L} &\sim \intdrei{p}
\Phi^{\Delta_0}(p) \frac{d}{d\Delta_0} \Phi^{\Delta_0}(p)\\
	& \sim \Delta_0^{-3/2}.
	\label{eq:dAdDelta0_Delta0}
\end{split}
\end{equation}
After integration, this gives rise to the expected infrared scaling behavior of
the amplitud
mode,\cite{Castellani1997a,Pistolesi2004,Strack2008,Obert2013} expressed in
terms of an external pairing field,
\begin{equation}
	A^{\Delta_0}(0) \sim \Delta_0^{-1/2}.
	\label{eq:A_Delta0}
\end{equation}
More generally, in dimension $2 < d < 4$ it reads $A^{\Delta_0}(0) \sim
\Delta_0^{(d-4)/2}$, which is similar to the behavior of the longitudinal
susceptibility in the non-linear sigma model in an external magnetic
field~\cite{Belitz2005} (here, the case $d = 2 + 1$ is relevant).

The leading contributions to the charge mode look similar to
Eq.~\eqref{eq:dA_dDelta0_TwoLoopLeading}, but with the anomalous
fermionic propagators replaced by normal ones,
\begin{equation}
\begin{split}
	\frac{d}{d\Delta_0}&C^{\Delta_0}(0)|_\text{2L} \sim -\intdrei{p}
\operatorname{Re} G(k+p)^2 \\ &\times \partial_{\Delta_0}
\bigl(A^{\Delta_0}(p)^2 +
\Phi^{\Delta_0}(p)^2\bigr) +\ldots
\end{split}
\end{equation}
The momentum factors resulting from the normal propagators weaken the
singularity of $\Phi^{\Delta_0}(p)^2$, so that the integral yields
$\tfrac{d}{d\Delta_0} C^{\Delta_0}(0)|_\text{2L} \sim \Delta_0^{-1/2}$ and thus
$C^{\Delta_0}(0)|_\text{2L} \sim \Delta_0^{1/2} + \operatorname{const}$. The
leading renormalization contribution to the anomalous (3+1) effective
interaction reads
\begin{equation}
\begin{split}
	\frac{d}{d\Delta_0}&X'^{\Delta_0}(0)|_\text{2L} \sim -\intdrei{p}
\operatorname{Re} G(k+p) F(k+p) \\ &\times \partial_{\Delta_0}
\bigl(\Phi^{\Delta_0}(p)^2 - A^{\Delta_0}(p)^2\bigr) +\ldots
\end{split}
\end{equation}
and is slightly more singular than $\tfrac{d}{d\Delta_0}
C^{\Delta_0}(0)|_\text{2L}$ due to the presence of one anomalous fermionic
propagator. A simple estimate hints at a logarithmic singularity of
$X'^{\Delta_0}(0)$, but its prefactor vanishes due to the approximate
particle-hole symmetry in the vicinity of the Fermi surface. In the magnetic
channel, the two-loop box diagrams yield a logarithmically singular contribution
to $\frac{d}{d\Delta_0}M^{\Delta_0}(0)|_\text{2L}$, which does not give rise to
singular behavior after integration. The two-loop contributions to the phase
mode are less singular than the propagator renormalization
diagrams. Note that simple estimates as above for propagator renormalization
diagrams with $\dot \Sigma$-insertions would yield a contribution to the phase
mode that diverges as $\Delta_0^{-3/2}$. Such a divergence would be
inconsistent with Eq.~\eqref{eq:ExPropAnsatz} and would lead to a drastic
violation of the $U(1)$ Ward identity. We did not detect such a contribution
numerically (see section~\ref{subsec:NumResults}), potentially due to 
cancellations caused by Ward identities.

The change in the infrared behavior of the vertex also impacts the
self-energy. The normal self-energy does not receive singular contributions, as
$C^{\Delta_0}(0)$ and $X'^{\Delta_0}(0)$ remain finite and the fluctuation
contributions are integrable in two dimensions at zero temperature. The
flow of the anomalous self-energy is altered due to the singular behavior of
$A^{\Delta_0}(0) \sim \Delta_0^{-1/2}$ to
\begin{equation}
\begin{split}
	\frac{d}{d\Delta_0} \Delta^{\Delta_0} &= - A^{\Delta_0}(0) \intdrei{p}
S_F^{\Delta_0}(p) + \ldots\\
	&\sim \Delta_0^{-1/2},
\end{split}
\end{equation}
where $S_F^{\Delta_0}$ is the anomalous component of the single-scale
propagator, which depends only weakly on $\Delta_0$. At one-loop level, one
obtains $\tfrac{d}{d\Delta_0} \Delta^{\Delta_0} \sim \mathcal O(1)$. The
anomalous self-energy thus becomes non-analytic at two-loop level as a function
of the external pairing field,
\begin{equation}
	\Delta^{\Delta_0} - \Delta^{\Delta_0 = 0} \sim \Delta_0^{1/2}.
	\label{eq:Delta_Delta0}
\end{equation}
More generally, this result reads $\Delta^{\Delta_0} - \Delta^{\Delta_0 = 0}
\sim \Delta_0^{(d-2)/2}$ in $2 < d < 4$ dimensions. Such a non-analytic
behavior is also found in the magnetization of the non-linear sigma model in
an external magnetic field.\cite{Belitz2005}

\subsection{Numerical results}
\label{subsec:NumResults}
We now present results for the effective interactions and the self-energy from
numerical solutions of the flow equations at two-loop level for a quarter-filled
system ($n = 1/2$), where the Fermi surface is almost circular. The results were
obtained by first integrating out the fermionic modes at $\Lambda > 0$ in the
presence of an external pairing field of the order of $\Delta_\text{MF} / 100$,
where $\Delta_\text{MF}$ is the mean-field gap, and subsequently reducing the
external pairing field in another flow.

\begin{figure}
	\centering
	\includegraphics[width=0.85\linewidth]{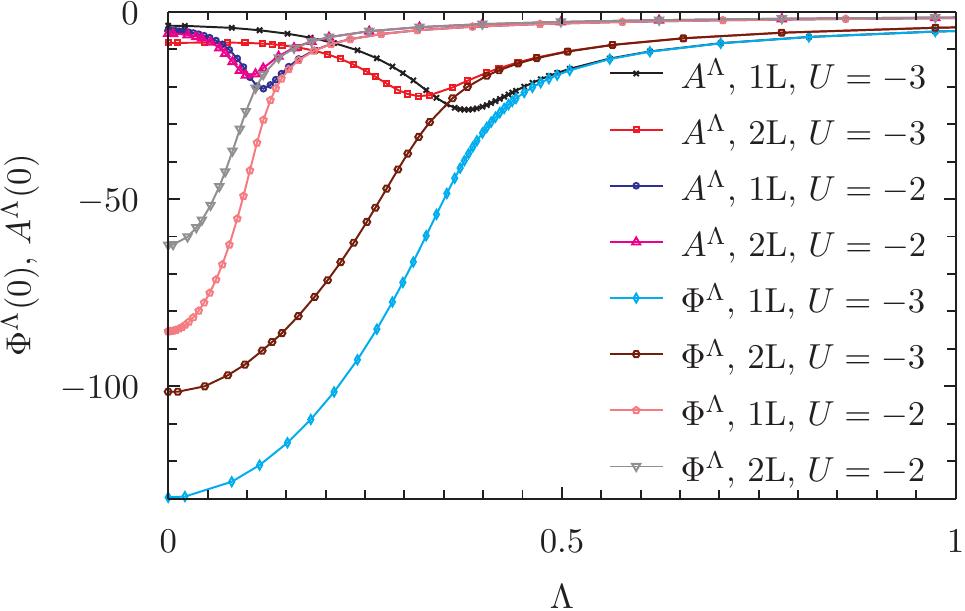}
	\caption{(Color online) Renormalization group flow of the amplitude
($A^\Lambda$) and phase ($\Phi^\Lambda$) coupling functions at fixed external
pairing field at one-loop (1L) and two-loop (2L) level for different
interactions $U$ for $n = 1/2$, $t' =-0.1$ and $\Delta_0 \approx
\Delta_\text{MF}/100$.}
	\label{fig9:APhi_OneTwoLoop}
\end{figure}
\begin{figure}
	\centering
	\includegraphics[width=0.85\linewidth]{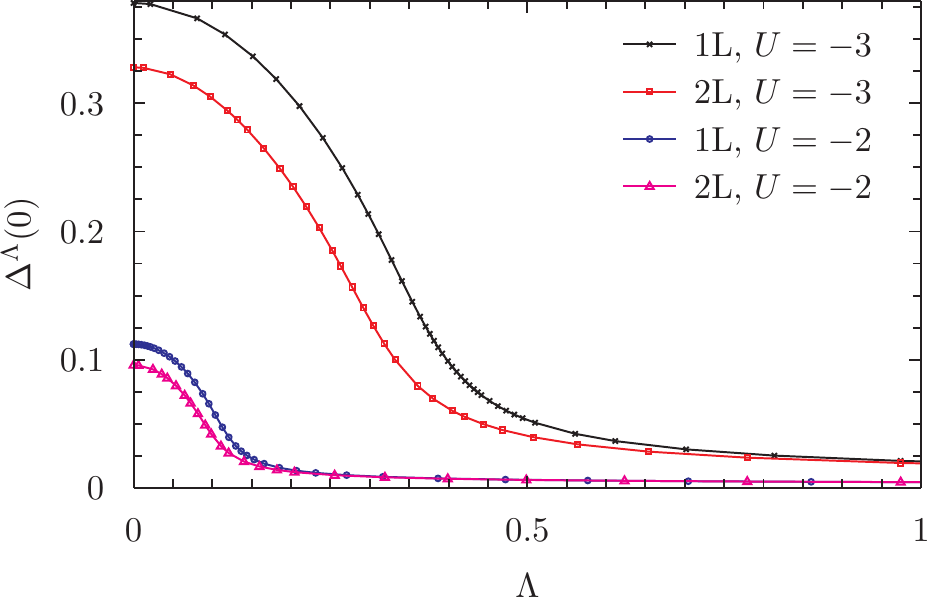}
	\caption{(Color online) Renormalization group flow of the superfluid gap at
fixed external pairing field at one-loop (1L) and two-loop (2L) level for
different interactions $U$ for $n = 1/2$, $t' = -0.1$ and $\Delta_0 \approx
\Delta_\text{MF}/100$.}
	\label{fig10:Delta_OneTwoLoop}
\end{figure}
In the presence of a not too small external pairing field, the flows at
two-loop level are qualitatively similar to those at one-loop level. This can
be seen in Fig.~\ref{fig9:APhi_OneTwoLoop}, which shows the flow of the
amplitude and phase mode of the gap at vanishing momentum and frequency,
$A^\Lambda(0)$ and $\Phi^\Lambda(0)$, for a quarter-filled ($n = 1/2$) system
with $t' = -0.1$ for different values of $U$. The critical scales $\Lambda_c$
(the scales where $|A^\Lambda(0)|$ is maximal) are, however, reduced due to
fluctuation corrections at two-loop level. The same observations can be made
for the flow of the anomalous self-energy at zero frequency, which is shown in
Fig.~\ref{fig10:Delta_OneTwoLoop} for the same parameters as in
Fig.~\ref{fig9:APhi_OneTwoLoop}. In the presence of a not too small 
external pairing field and for $|U| \leq 3$, the $\Lambda$-dependence
of the anomalous self-energy is still qualitatively similar to that
in mean-field theory, $\Delta^\Lambda \approx \sqrt{\Lambda_c^2 -
\Lambda^2}$ for $\Lambda < \Lambda_c$,  even at two-loop level. This means that
the critical scale and gap are mainly reduced by fluctuations \emph{above} 
$\Lambda_c$. For smaller external pairing fields or larger values of $|U|$, the 
agreement worsens because the gap at $\Lambda = 0$ gets somewhat larger than 
expected from the above relation. This indicates an increasing impact of phase 
fluctuations and is accompanied by a change in the behavior of $A^\Lambda(0)$ 
for small $\Lambda$. Instead of a monotonic decrease in absolute value below the 
critical scale as shown in Fig.~\ref{fig9:APhi_OneTwoLoop}, $A^\Lambda(0)$ first 
decreases in absolute value below the critical scale and then slightly increases 
at low scales. For $\Delta_0 \approx \Delta_\text{MF} / 100$, this effect was 
either absent ($|U| \leq 3$) or small, so that long-range phase fluctuations 
were mostly treated in the pairing field flows.

The momentum and frequency dependence of the self-energy and exchange
propagators is qualitatively similar to the one-loop
approximation.\cite{Eberlein2013} The same holds for the flows of the effective
interactions in the magnetic, charge and anomalous (3+1) channels. We therefore
do not show results for these quantities.

\begin{figure}
	\centering
	\includegraphics[width=0.85\linewidth]{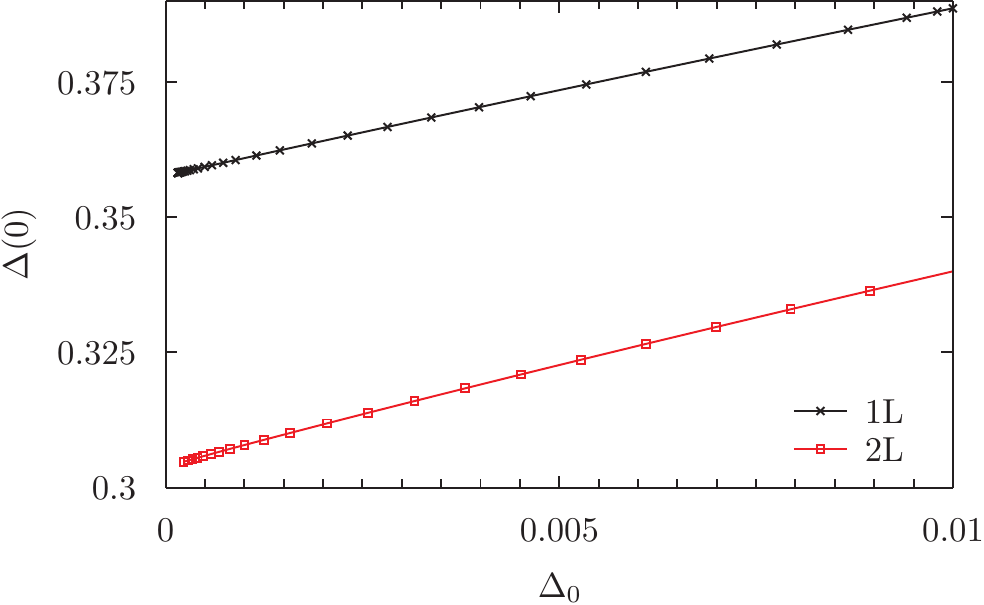}
	\caption{(Color online) Pairing field flow of the anomalous self-energy at
one-loop (1L) and two-loop (2L) level for $U = -3$, $t' = -0.1$
and $n = 1/2$. The external pairing field $\Delta_0$ is used as the flow
parameter and was chosen as $\Delta_0 = \Delta_\text{MF} / 50$ in the fermionic
flow.}
	\label{fig11:Delta_Delta0_OneTwoLoop}
\end{figure}

\begin{figure}
	\centering
	\includegraphics[width=0.85\linewidth]{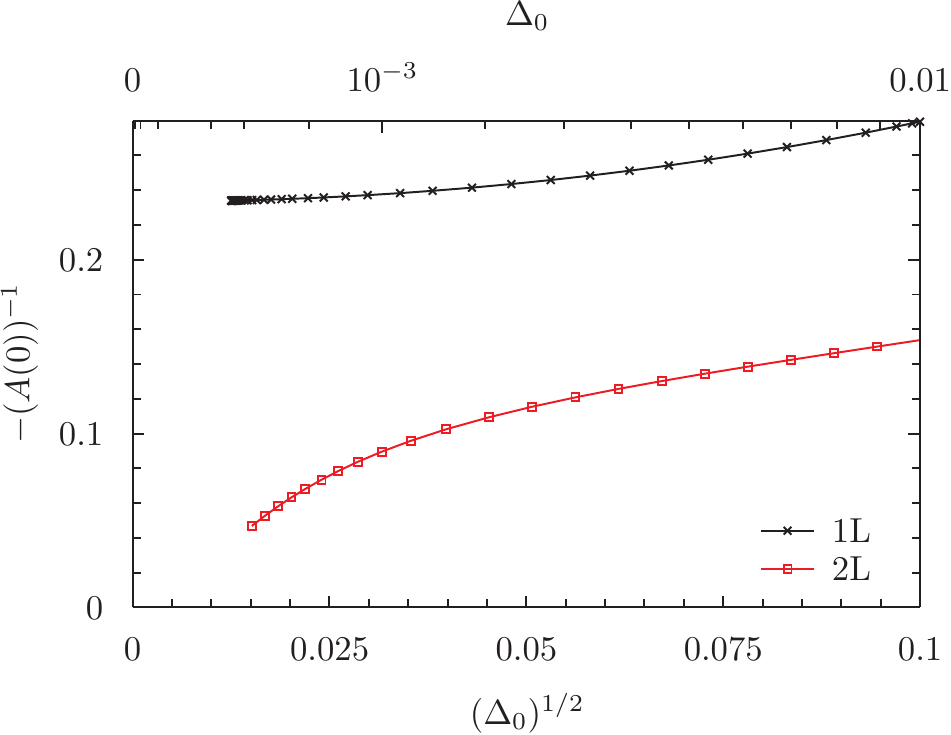}
	\caption{(Color online) Pairing field flow of the amplitude coupling function
$A$ at one-loop (1L) and two-loop (2L) level for the same parameters as in
Fig.~\ref{fig11:Delta_Delta0_OneTwoLoop}.}
	\label{fig12:A_Delta0_OneTwoLoop}
\end{figure}

\begin{figure}
	\centering
	\includegraphics[width=0.85\linewidth]{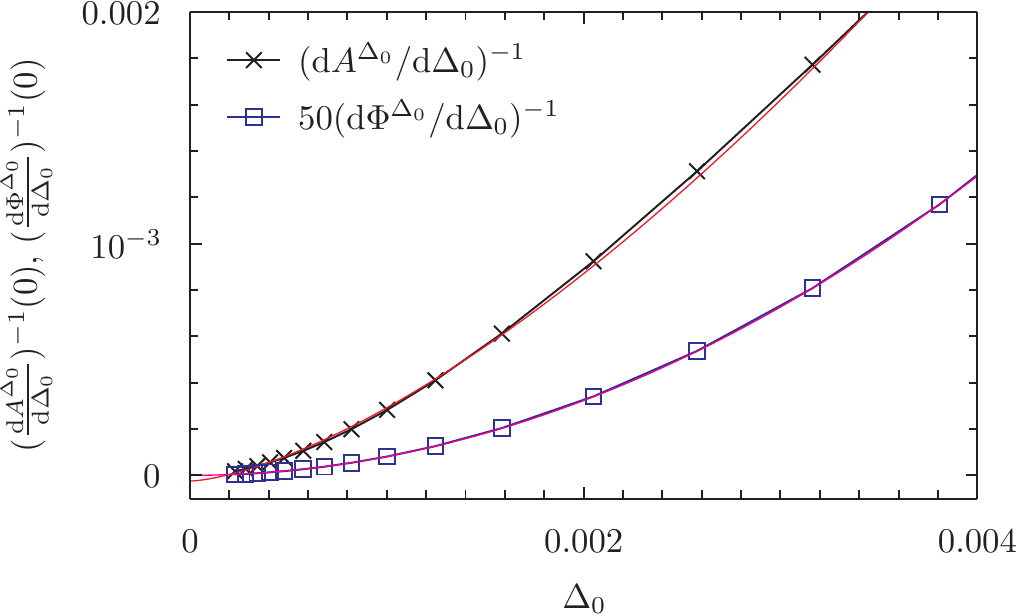}
	\caption{(Color online) Pairing field flow of the inverse of the derivatives
of the amplitude $A$ and phase $\Phi$ coupling functions with respect to the
external pairing field $\Delta_0$ at two-loop level for the same parameters as
in Fig.~\ref{fig11:Delta_Delta0_OneTwoLoop}.
The red and magenta lines are fits to the expected dependence on the external
pairing field (see text). For a better presentation, the result for the phase
mode is rescaled by factors of $50$.}
	\label{fig13:dAdD_dPhidD_inverse_TwoLoop}
\end{figure}

The impact of phase fluctuations can be studied in a controlled way by
eliminating the external pairing field in a second flow. The results of such
pairing field flows are shown in
Figs.~\ref{fig11:Delta_Delta0_OneTwoLoop},~\ref{fig12:A_Delta0_OneTwoLoop}
and~\ref{fig13:dAdD_dPhidD_inverse_TwoLoop}.
Figure~\ref{fig11:Delta_Delta0_OneTwoLoop} shows the reduction of the anomalous
self-energy in pairing field flows at one- and two-loop level for $U
= -3$, $t' = -0.1$ and $n = 1/2$, which is mainly caused by amplitude and
long-range phase fluctuations. Depending on the initial size of the
external pairing field, the anomalous self-energy is reduced around 10\,\% in
the pairing field flow, with a slightly stronger reduction for larger initial
external pairing fields. The $\Delta_0$-dependence of the gap at one- and
two-loop level is linear to a very good approximation. This is expected at
one-loop level. At two-loop level, the numerically accessible external pairing
fields are too large for resolving the expected non-analytic behavior on the
scale of Fig.~\ref{fig11:Delta_Delta0_OneTwoLoop}. Fitting $\Delta^{\Delta_0} =
a + b \Delta_0^{1/2} + c \Delta_0$ to the $\Delta_0$-dependence of the gap at 
two-loop level yields only a small coefficient for the term $\sim
\Delta_0^{1/2}$.

The amplitude mode $A^{\Delta_0}$ gets strongly renormalized during the pairing
field flow as can be seen in Fig.~\ref{fig12:A_Delta0_OneTwoLoop}, which
compares the one- and two-loop approximations for a quarter-filled system with
$U = -3$ and $t' = -0.1$. In both cases, the flows for $\Lambda > 0$ were 
computed in the presence of an external pairing field $\Delta_0 =
\Delta_\text{MF} / 50$, which was reduced by a factor $\approx 50$ in the
pairing field flow. Smaller external pairing fields were not accessible within
our framework of approximations due to remnants of the violation of the Ward
identity for the global $U(1)$ charge symmetry that cannot be cured with the 
abovementioned simple projection method for enforcing the Ward
identity.\footnote{As a consequence of the violation of the Ward identity for 
the $U(1)$ charge symmetry, the frequency dependence of $\Phi(q)$ at $\Lambda = 
0$ deviates slightly from the quadratic dependence in 
Eq.~\eqref{eq:ExPropAnsatz} at low frequencies, yielding a small plateau. This 
cannot be cured with our simple projection scheme for enforcing the Ward 
identity. At very small external pairing fields this plateau leads to an 
overestimation of phase fluctuations, which also gives rise to the small offset 
seen in Fig.~\ref{fig13:dAdD_dPhidD_inverse_TwoLoop} for
$(\text{d}A^{\Delta_0}/\text{d}\Delta_0)^{-1}$.} The results in 
Fig.~\ref{fig12:A_Delta0_OneTwoLoop} are plotted
in such a way that the scaling in Eq.~\eqref{eq:A_Delta0} would yield a linear
dependence near $\Delta_0 = 0$. Fluctuations at two-loop level clearly lead
to a strong renormalization of the amplitude mode and tend to suppress
$(-A(0))^{-1}$ towards zero. However, due to the limited range of
accessible pairing fields, the behavior in the limit $\Delta_0 \rightarrow 0$ is
not apparent from this plot. From the numerical data, we also cannot draw
conclusions on the behavior of the effective interactions in the particle-hole
channel in this limit for the same reason.
Figure~\ref{fig13:dAdD_dPhidD_inverse_TwoLoop} shows the inverse of the
derivatives of the amplitude and phase coupling functions at $q = 0$ with
respect to the external pairing field as computed during the same pairing field
flow as in Fig.~\ref{fig12:A_Delta0_OneTwoLoop}. 
From Eq.~\eqref{eq:ExPropAnsatz} and~\eqref{eq:dAdDelta0_Delta0} one expects
$(\text{d}\Phi/\text{d}\Delta_0)(0) \propto \Delta_0^{-2}$ and
$(\text{d}A/\text{d}\Delta_0)(0) \propto \Delta_0^{-3/2}$. In 
Fig.~\ref{fig13:dAdD_dPhidD_inverse_TwoLoop}, the numerical results are 
compared to fits using the function $f(\Delta_0) = a + b \Delta_0^n$, where $n = 
2$ for $(\text{d}\Phi/\text{d}\Delta_0)^{-1}$ and $n = 3/2$ for 
$(\text{d}A/\text{d}\Delta_0)^{-1}$, showing good agreement with the 
expected dependence on $\Delta_0$. The small offsets near $\Delta_0 = 0$ are 
remnants of the violation of the Ward identity for the global $U(1)$ charge 
symmetry.

\begin{figure}
	\centering
	\includegraphics[width=0.85\linewidth]{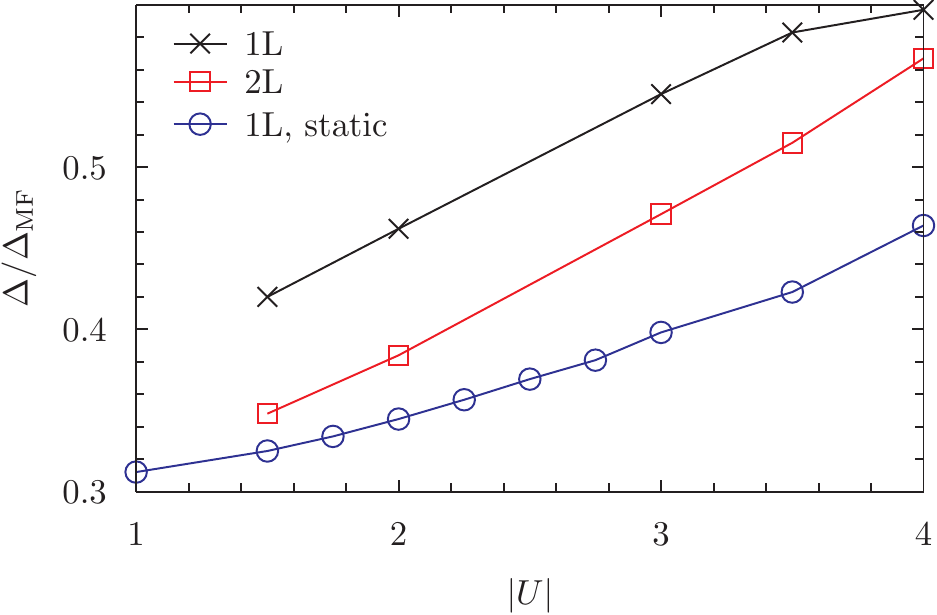}
	\caption{(Color online) Gap ratio $\Delta / \Delta_\text{MF}$ as a function
of $U$ from one-loop (1L) and two-loop (2L) flows for $n = 1/2$ and $t' =
-0.1$. Also shown are results from a static approximation (1L, static) in which
all frequency dependences are neglected.}
	\label{fig14:Gapreduction_n5}
\end{figure}
Figures~\ref{fig9:APhi_OneTwoLoop} and~\ref{fig10:Delta_OneTwoLoop} already
gave an impression of the renormalization of the critical scale and the gap by 
fluctuations at two-loop level. The impact of these fluctuations also
depends on the size of $U$. This can be seen in
Fig.~\ref{fig14:Gapreduction_n5}, which shows the gap ratio $\Delta /
\Delta_\text{MF}$ as a function of the interaction for a quarter-filled system
with $t' = -0.1$. $\Delta = \Delta(0)$ is the gap as obtained
from extrapolating pairing field flows to the limit $\Delta_0 \rightarrow 0$ and
$\Delta_\text{MF}$ is the gap in mean-field approximation. The one-loop results
are in very good agreement with those of Ref.~\onlinecite{Eberlein2013} despite
the differences in the approximations for the momentum and frequency dependence
of exchange propagators. The observed increase of $\Delta /
\Delta_\text{MF}$ with $|U|$ is consistent with the behavior in the limits $U
\rightarrow 0^-$ and $U \rightarrow -\infty$. These limits are accessible in
perturbation theory~\cite{Georges1991,Martin-Rodero1992} or by mapping the
attractive Hubbard model at finite doping to the Heisenberg model in a magnetic
field, respectively. Results for the staggered magnetization in the latter were
obtained numerically in Ref.~\onlinecite{Luescher2009}. In the coupling range
considered, the gaps at two-loop level are 5 - 20 \% smaller than at one-loop
level.  For smaller values of $|U|$, it is difficult to compute the gap from a
numerical solution of the flow equations, because the gap and the critical scale
decrease exponentially. It is expected that the two-loop result approaches the
one-loop result for smaller values of $U$, because the phase space for
fluctuations at two-loop level decreases with $|U|$. It is interesting that the
gap ratios at one- and two-loop level approach each other with increasing $|U|$.
This may indicate that the Katanin scheme overestimates certain fluctuation
contributions, which are compensated by two-loop contributions with overlapping
loops at larger $|U|$. Figure~\ref{fig14:Gapreduction_n5} also shows results
from a static one-loop approximation, in which the frequency dependence of the
self-energy and vertex is neglected. Note that the gaps from this approximation
are even smaller than those from the two-loop approximation, indicating that the
former overestimates the impact of fluctuations when using the frequency
regulator in Eq.~\eqref{eq:Regulator}.

\section{Summary}
\label{sec:Summary}
We have analyzed flow equations for the two-particle vertex in the fermionic
functional renormalization group at two-loop level and reformulated them
effectively as one-loop equations. In two-loop contributions with overlapping
loops, the insertion of two vertices that are connected by a full and a
single-scale propagator can be reexpressed through the one-loop result for the
scale-derivative of the vertex. This is similar in spirit to the replacement of
tadpole insertions by scale-derivatives of the self-energy in the Katanin
scheme. The reformulation is exact to the third order in the effective
interaction, sheds light on the physics described by the two-loop
renormalization contributions, and allows for their efficient numerical
treatment.

The proposed scheme is based on a decomposition of the vertex in charge, 
magnetic and pairing channels. Using this decomposition, the singular 
dependence of the vertex on momenta and frequencies can be described within a 
reasonable numerical effort also at two-loop level. The scheme allows to 
continue renormalization group flows into phases with broken symmetries, which 
we demonstrated for the superfluid ground state of the attractive Hubbard model.

Using simple estimates for the most singular diagrams, we analyzed the infrared
behavior of the vertex and the self-energy in the ground state of an $s$-wave
superfluid in the BCS regime. We find that the two-loop scheme captures the
expected singular behavior of the amplitude mode as well as the non-analytic
behavior of the order parameter in the limit where the external pairing field
vanishes. In a description using auxiliary bosons for the order parameter,
this infrared behavior is governed by a non-Gaussian fixed
point.\cite{Pistolesi2004,Strack2008} Thus, our approach captures non-Gaussian
fluctuations, although the related fixed point structure is less transparent
than in the partially bosonized approach.

We argue that the vertex in the ground state of a fermionic $s$-wave superfluid
in two dimensions does not exhibit infrared singularities beyond those in the
Cooper channel that are already known from the singular infrared behavior of
interacting bosons. Our formalism yields a unified description of the reduction
of the anomalous self-energy by particle-hole and collective fluctuations. In
comparison to the one-loop approximation, the obtained superfluid order
parameters at two-loop order are slightly smaller.

The formalism presented in this article may be useful also in other contexts. As
it treats all interaction channels on equal footing and captures single-particle
as well as collective fluctuations, it might be a convenient tool for the study
of competing orders in systems of correlated fermions, like the repulsive
Hubbard model, possibly in conjunction with mean-field theory for the low-energy
modes below the scale for symmetry breaking.\cite{JWang2014} Quite generally, 
it may be helpful for applying the fermionic functional renormalization group at 
larger interactions.

\begin{acknowledgments}
I would like to thank N.~Hasselmann, T.~Holder, C.~Honerkamp,
C.~Husemann, S.~Maier, W.~Metzner, B.~Obert, M.~Salmhofer and K.~Veschgini for
valuable discussions.
\end{acknowledgments}

%

\end{document}